\documentclass[onefignum,onetabnum]{siamart171218}

\usepackage{braket,amsfonts}
\usepackage{array}
\usepackage[caption=false]{subfig}
\usepackage{pgfplots,bm,multirow}

\newsiamthm{claim}{Claim}
\newsiamremark{remark}{Remark}
\newsiamremark{hypothesis}{Hypothesis}
\crefname{hypothesis}{Hypothesis}{Hypotheses}

\usepackage{algorithmic}

\usepackage{graphicx,epstopdf}

\Crefname{ALC@unique}{Line}{Lines}

\usepackage{amsopn}

\usepackage{xspace}
\usepackage{bold-extra}
\usepackage[most]{tcolorbox}

\colorlet{texcscolor}{blue!50!black}
\colorlet{texemcolor}{red!70!black}
\colorlet{texpreamble}{red!70!black}
\colorlet{codebackground}{black!25!white!25}

\lstdefinestyle{siamlatex}{%
  style=tcblatex,
  texcsstyle=*\color{texcscolor},
  texcsstyle=[2]\color{texemcolor},
  keywordstyle=[2]\color{texemcolor},
  moretexcs={cref,Cref,maketitle,mathcal,text,headers,email,url},
}

\tcbset{%
  colframe=black!75!white!75,
  coltitle=white,
  colback=codebackground, 
  colbacklower=white, 
  fonttitle=\bfseries,
  arc=0pt,outer arc=0pt,
  top=1pt,bottom=1pt,left=1mm,right=1mm,middle=1mm,boxsep=1mm,
  leftrule=0.3mm,rightrule=0.3mm,toprule=0.3mm,bottomrule=0.3mm,
  listing options={style=siamlatex}
}

\newtcblisting[use counter=example]{example}[2][]{%
  title={Example~\thetcbcounter: #2},#1}

\newtcbinputlisting[use counter=example]{\examplefile}[3][]{%
  title={Example~\thetcbcounter: #2},listing file={#3},#1}

\DeclareTotalTCBox{\code}{ v O{} }
{ 
  fontupper=\ttfamily\color{black},
  nobeforeafter,
  tcbox raise base,
  colback=codebackground,colframe=white,
  top=0pt,bottom=0pt,left=0mm,right=0mm,
  leftrule=0pt,rightrule=0pt,toprule=0mm,bottomrule=0mm,
  boxsep=0.5mm,
  #2}{#1}

\patchcmd\newpage{\vfil}{}{}{}
\begin{tcbverbatimwrite}{tmp_\jobname_header.tex}
\title{Fast convergence and asymptotic preserving of the General Synthetic Iterative Scheme \thanks{Submitted to the editors DATE.
\funding{Financial supports from the Engineering and Physical Sciences Research Council (UK) under grant EP/R041938/1 and European Union’s Horizon 2020 Research and Innovation Programme under the Marie Skłodowska-Curie grant agreement number 793007 are acknowledged.}
}}

\author{
	Wei Su\thanks{James Weir Fluids Laboratory, Department of Mechanical and Aerospace Engineering, University of Strathclyde, Glasgow G1 1XJ, UK}
	\and Lianhua Zhu\footnotemark[2] 
	\and Lei Wu\thanks{Department of Mechanics and Aerospace Engineering, Southern University of Science and Technology, Shenzhen 518055, China (\email{wul@sustech.edu.cn}) }
}

\headers{GSIS: Fast convergence and asymptotic  preserving}{Wei Su, Lianhua Zhu, Lei Wu}
\end{tcbverbatimwrite}
\title{Fast convergence and asymptotic preserving of the General Synthetic Iterative Scheme \thanks{Submitted to the editors DATE.
\funding{Financial supports from the Engineering and Physical Sciences Research Council (UK) under grant EP/R041938/1 and European Union’s Horizon 2020 Research and Innovation Programme under the Marie Skłodowska-Curie grant agreement number 793007 are acknowledged.}
}}

\author{
	Wei Su\thanks{James Weir Fluids Laboratory, Department of Mechanical and Aerospace Engineering, University of Strathclyde, Glasgow G1 1XJ, UK}
	\and Lianhua Zhu\footnotemark[2]
	\and Lei Wu\thanks{Department of Mechanics and Aerospace Engineering, Southern University of Science and Technology, Shenzhen 518055, China (\email{wul@sustech.edu.cn}) }
}

\headers{GSIS: Fast convergence and asymptotic  preserving}{Wei Su, Lianhua Zhu, Lei Wu}

\ifpdf
\hypersetup{ pdftitle={Guide to Using  SIAM'S \LaTeX\ Style} }
\fi

\definecolor{mygreen}{rgb}{0.0,0.55,0.55}

\begin{document}
\maketitle

\begin{tcbverbatimwrite}{tmp_\jobname_abstract.tex}
\begin{abstract}
Recently the general synthetic iteration scheme (GSIS) is proposed to find the steady-state solution of the Boltzmann equation~\cite{SuArXiv2019}, where various numerical simulations have shown that (i) the steady-state solution can be found within dozens of iterations at any Knudsen number $K$, and (ii) the solution is accurate even when the spatial cell size in the bulk region is much larger than the molecular mean free path, i.e. Navier-Stokes solutions are recovered at coarse grids. The first property indicates that the error decay rate between two consecutive iterations decreases to zero with $K$, while the second one implies that the GSIS is asymptotically preserving the Navier-Stokes limit. This paper is dedicated to the rigorous proof of both properties. %
\end{abstract}

\begin{keywords}
  gas kinetic equation, convergence rate, asymptotic preserving
\end{keywords}

\begin{AMS}
	76P05, 
	65L04, 
  65M12 
 
\end{AMS}
\end{tcbverbatimwrite}
\begin{abstract}
Recently the general synthetic iteration scheme (GSIS) is proposed to find the steady-state solution of the Boltzmann equation~\cite{SuArXiv2019}, where various numerical simulations have shown that (i) the steady-state solution can be found within dozens of iterations at any Knudsen number $K$, and (ii) the solution is accurate even when the spatial cell size in the bulk region is much larger than the molecular mean free path, i.e. Navier-Stokes solutions are recovered at coarse grids. The first property indicates that the error decay rate between two consecutive iterations decreases to zero with $K$, while the second one implies that the GSIS is asymptotically preserving the Navier-Stokes limit. This paper is dedicated to the rigorous proof of both properties. %
\end{abstract}

\begin{keywords}
  gas kinetic equation, convergence rate, asymptotic preserving
\end{keywords}

\begin{AMS}
	76P05, 
	65L04, 
  65M12 

\end{AMS}



\section{Introduction}
Rarefied gas flows are encountered in many engineering problems, including space vehicles' reentry,  micro-electromechanical systems, and shale gas extraction. Although several macroscopic equations are proposed~\cite{CE,Grad1949,Sone2002Book}, they only work when the Knudsen number $K$, defined as the ratio of the molecular mean free path to the characteristic system length, is small. To describe the rarefied gas dynamics from the continuum to free molecular flow regimes, the Boltzmann equation from the gas kinetic theory is needed. However, it is a grand challenge to solve the Boltzmann equation, as the numerical scheme should be carefully designed, especially when $K$ is small so that the Boltzmann collision operator becomes stiff. For example, in the direct simulation Monte Carlo method~\cite{Bird1994}, it is required that the numerical scale (i.e. the spatial cell size $\Delta{x}$ and time step $\Delta{t}$) is smaller than the corresponding kinetic scale (i.e. the mean free path and mean collision time of gas molecules), otherwise the numerical dissipation will lead to inaccurate solutions. This would become prohibitive in the multiscale simulation involving the near-continuum flow where the Knudsen number is small.

Many efforts have been devoted to solving the Boltzmann equation or its simplified model equations under a numerical scale larger than the kinetic one~\cite{Luc2000JCP,Filbet2010JCP,Filbet2011,Dimarco2017Siam,Ju2017JSC,Xu2010JCP,guo2013discrete}. The common feature of these numerical methods is that they are time-dependent. Some schemes asymptotically preserve the Euler limit, i.e. they become a consistent discretization of Euler equations when $K$ goes to zero~\cite{Filbet2010JCP,Filbet2011}, while some asymptotically preserve the Navier-Stokes limit when $\Delta{t}$ is much larger than the mean collision time~\cite{Dimarco2017Siam,Ju2017JSC}, or when both $\Delta{t}$ and $\Delta{x}$ are much larger than the corresponding kinetic scales~\cite{Xu2010JCP,guo2013discrete,Guo2019UP_arXiv}.

It should be noted that in continuum flows the turbulence is common and time-dependent numerical solvers for Navier-Stokes equations are highly demanded. However, in rarefied gas dynamics the turbulence is absent and steady-state solutions are of particular interest. Therefore, the implicit solver is strongly recommended, especially in multiscale problems where the Knudsen layer (a few mean free path away from the solid surface) must be resolved (see the thermal flow in section~\ref{thermal_edge} for an example). In designing the implicit solver, one has to consider the following two major problems: (i) how to find the steady-state solution quickly and (ii) how to recover the solution of Navier-Stokes equations in the continuum flow regime, on coarse spatial grids that are adequate to capture the hydrodynamics but much larger than the mean free path?

A simple way to obtain the steady-state solution of gas kinetic equations is to use the conventional iterative scheme (CIS), where the collision operator is splitted into the gain term and the loss term. The streaming operator and loss term are calculated at the current iteration step, while the gain term is evaluated at the previous iteration step. Therefore, the resultant ordinary differential equation can be effectively solved by the sweep technique after discretization in physical space, without involving the inverse of large-size matrix~\cite{Ho2019CPC}. The CIS is efficient and accurate when the Knudsen number is large, where converged solution can be obtained within a few iterations. However, the iteration number increases significantly when $K$ is small. Moreover, the CIS does not asymptotically preserve the Navier-Stokes limit, so  a huge number of spatial cells should be used to capture the flow dynamics when $K$ is small~\cite{WANG201833}.

It is noted that these two major deficiencies of CIS not only appear in gas system, but also in other kinetic systems for neutron transport, thermal radiation, and phonon transport in crystals, to name just a few. The exciting breakthrough to tackle these deficiencies is firstly achieved probably in neutron transport, where the diffusion synthetic acceleration (DSA) is developed, see the review~\cite{DSA2002}. In  DSA, the mesoscopic kinetic equation is solved together with a diffusion equation for the macroscopic variable appearing in the gain part of the collision operator. This diffusion equation is exactly the limiting equation derived from the kinetic equation in optical thick regions (or equivalently when $K$ is small)~\cite{LARSEN1987283}. Not only fast convergence to steady state but also asymptotic preserving are achieved on coarse physical grids, in the whole range of Knudsen number.

The DSA scheme has been extended to some special rarefied gas flows, such as the Poiseuille~\cite{Valougeorgis2003,LeiJCP2017} and Couette~\cite{SU2019573} flows, where the flow velocity appearing in the collision operator is perpendicular to the computational domain, thus its macroscopic governing equation is an inhomogeneous diffusion equation. In the continuum flow regime the inhomogeneous term vanishes and the macroscopic equation is reduced to the diffusion equation, while in rarefied regimes the inhomogeneous term dominates and captures the rarefaction effects exactly. However, for general rarefied gas flows, the limiting macroscopic equations are not the diffusion equation but the Navier-Stokes equation as according to the Chapman-Enskog expansion~\cite{CE}. Thus, in a recent paper we proposed the general synthetic iterative scheme (GSIS), where the macroscopic synthetic equations take the form of linearized Navier-Stokes equations but with source terms describing high-order rarefaction effects~\cite{SuArXiv2019}.

Since the synthetic equations contain the continuity equation (obtained from mass conservation) that is not of diffusion type, GSIS may have different behaviors when compared to other DSA schemes. Although the numerical examples in Ref.~\cite{SuArXiv2019} have shown that GSIS has the properties of fast convergence to steady state and asymptotic Navier-Stokes preserving, the mathematical prove is lacking. It is therefore the aim of this paper to rigorously investigate the two properties of GSIS.

The remainder of this paper is organized as follows. In sections~\ref{CIS_convergence} and~\ref{GSIS:section} the convergence rates of CIS and GSIS for the linearized BGK kinetic equation~\cite{Bhatnagar1954} are calculated, respectively. The reason of explicitly including Newton's law of shear stress and Fourier's law of heat conduction in the macroscopic synthetic equations is analyzed. In section~\ref{AP_property} the asymptotic preserving property of GSIS is studied and it is found that GSIS preserves the Navier-Stokes limit with $\Delta{x}\sim{}O(1)$, as long as this spatial resolution is able to capture the hydrodynamic behavior. In section~\ref{sec_num}, several numerical simulations, including the most challenging simulation of thermally-induced flow in the near-continuum regime, are preformed to show the fast convergence and asymptotic Navier-Stokes preserving properties of the GSIS. 
A summary is given in section~\ref{sec:conclusion}.

\section{The convergence rate of CIS and false convergence}\label{CIS_convergence}

To make the mathematical derivation simple but keep the essential flow physics, we consider the linearized BGK equation in general two-dimensional problems:
\begin{equation}\label{LBE_BGK}
\bm{v}\cdot\frac{\partial{h}(\bm{x},\bm{v})}{\partial\bm{x}}=\frac{h_{eq}(\bm{x},\bm{v})-h(\bm{x},\bm{v})}{K},
\end{equation}
where $h(\bm{x},\bm{v})$ is the velocity distribution function at location $\bm{x}=(x_1,x_2)$, $\bm{v}=(v_1,v_2)$ is the molecular velocity, $K$ is the Knudsen number, and $h_{eq}$ is the reference velocity distribution function:
\begin{equation}\label{LBE_BGK_gain}
h_{eq}(\bm{x},\bm{v})=\varrho(\bm{x})+2\bm{u}(\bm{x})\cdot\bm{v}+\tau(\bm{x})\left(v^2-1\right),
\end{equation}
with $\varrho$, $\bm{u}=(u_1,u_2)$ and $\tau$ being the density, flow velocity and temperature deviated from the global equilibrium state, respectively. They are related to the velocity distribution function $h$ as 
\begin{eqnarray}\label{Macro0}
M(\bm{x})=[\varrho,u_1,u_2,\tau]=\int{m}{h(\bm{x},\bm{v})}E(\bm{v}) d\bm{v},
\end{eqnarray}
where ${E}(\bm{v})=\exp(-v^2)/\pi$ is the velocity distribution function at global equilibrium and $m=[1,v_1, v_2, v^2-1]$ are the collision invariants. Also, the shear stress $\sigma_{ij}$ (with $i,j=1,2$) and heat flux $q_i$ are defined as
\begin{eqnarray}
\sigma_{ij}(\bm{x})=2\int \left(v_iv_j-\frac{v^2}{2}\delta_{ij}\right)h(\bm{x},\bm{v})E(\bm{v})d\bm{v}, \label{NS_shear_def}\\
q_i(\bm{x})=\int {v_i}\left(v^2-2\right)h(\bm{x},\bm{v})E(\bm{v})d\bm{v}, \label{NS_heat_def}
\end{eqnarray}
with $\delta_{ij}$ the Kronecker delta.




In CIS, given the value of velocity distribution function $h^{(k)}(\bm{x},\bm{v})$ at the $k$-th iteration step, its value at the next iteration step is calculated by solving the following equation:
\begin{equation}\label{LBE_iteration}
h^{(k+1)}+
K\bm{v}\cdot\frac{\partial
	{h}^{(k+1)}}{\partial{\bm{x}}}=h_{eq}(h^{(k)}),
\end{equation}
and the process is repeated until the maximum relative difference between successive estimates of macroscopic quantities, i.e. 
\begin{equation}\label{general_epsilon}
\epsilon=max\left\{\sqrt{\frac{\int|M^{(k+1)}-M^{(k)}|^2\mathrm{d}\bm{x}}{\int|M^{(k)}|^2\mathrm{d}\bm{x}}}\right\},
\end{equation}
is less than a preassigned value.

Although in practical numerical simulations the streaming operator $\partial/\partial{\bm{x}}$ is discretized, e.g. by the finite-difference, finite-volume, or discontinuous Galerkin (DG) methods~\cite{WeiSuJCP1,Su2019IDG}, here it is kept intact when calculating the convergence rate of iteration; that of the discretized version of Eq.~\eqref{LBE_iteration} will be tested in numerical simulations in section~\ref{sec_num} below.  

To calculate the convergence rate, we first define the following error function between velocity distribution functions at two consecutive iteration steps:
\begin{eqnarray}
Y^{(k+1)}(\bm{x},\bm{v})=h^{(k+1)}(\bm{x},\bm{v})-h^{(k)}(\bm{x},\bm{v}), \label{Y_expression}
\end{eqnarray}
where according to Eq.~\eqref{LBE_iteration} the error function $Y^{(k+1)}(\bm{x},\bm{v})$ is found to satisfy
\begin{equation}\label{LBE_iteration2}
Y^{(k+1)}+
K\bm{v}\cdot\frac{\partial
	{Y}^{(k+1)}}{\partial{\bm{x}}}=\Phi_\varrho^{(k)}+2\Phi_{u_1}^{(k)}v_1+2\Phi_{u_2}^{(k)}v_2+\Phi_\tau^{(k)}\left(v^2-1\right),
\end{equation}
with the following error functions for macroscopic quantities between two consecutive iteration steps: 
\begin{eqnarray}
\Phi^{(k+1)}_M(\bm{x})=M^{(k+1)}(\bm{x})-M^{(k)}(\bm{x})={\int{}mY^{(k+1)}(\bm{x},\bm{v})E(\bm{v})d\bm{v}}. \label{Phi_expression}
\end{eqnarray}

Second, to determine the convergence rate $\omega$ we perform the Fourier stability analysis by seeking the eigenvalue $\omega$ and eigenfunctions $y(\bm{v})$ and $\alpha_M=[\alpha_\varrho,\alpha_{u_1}, \alpha_{u_2}, \alpha_\tau]$  of the following forms:
\begin{eqnarray}
Y^{(k+1)}(\bm{x},\bm{v})=\omega^{k}y(\bm{v})\exp(i\bm{\theta}\cdot{\bm{x}}),\label{Y_ansatz} \\
\Phi^{(k+1)}_{M}(\bm{x})=\omega^{k+1}\alpha_M\exp(i\bm{\theta}\cdot{\bm{x}}), \label{Phi_ansatz}
\end{eqnarray}
where $i$ is the imaginary unit and $\bm{\theta}=(\theta_1,\theta_2)$ is the wave vector of perturbance. It should be noted that the factor $\omega^k$ emerges in Eq.~\eqref{Y_ansatz} rather than the factor $\omega^{k+1}$ in Eq.~\eqref{Phi_ansatz} because from Eq.~\eqref{LBE_iteration2} we know $Y^{(k+1)}$ is
determined by macroscopic quantities in the $k$-th iteration step.

Substituting Eqs.~\eqref{Y_ansatz} and~\eqref{Phi_ansatz} into Eqs.~\eqref{LBE_iteration2} and~\eqref{Phi_expression} we obtain the following expression for $y(\bm{v})$:
\begin{equation}\label{y_solution_CIS}
y(\bm{v})=\frac{\alpha_\varrho+2\alpha_{u_1}v_1+2\alpha_{u_2}v_2+\alpha_\tau(v^2-1)}{ 1+iK\bm{\theta}\cdot\bm{v} }.
\end{equation}
In the following we assume the wave vector of perturbation satisfies $|\bm{\theta}|^2=\theta_1^2+\theta_2^2=1$. Although in reality the perturbation may have various values of $\bm{\theta}$, their corresponding convergence rates do not interact because the kinetic equation is linear. Moreover from the denominator in Eq.~\eqref{y_solution_CIS} we see that the convergence rate depends only on the product of $K\bm{\theta}$. If $|\bm{\theta}|\neq1$, the convergence rate at specific values of $\bm{\theta}$ and $K$ can be calculated by replacing the Knudsen number with $K\bm{\theta}$.

Finally, with Eqs.~\eqref{Phi_expression}, \eqref{Y_ansatz}, \eqref{Phi_ansatz} and~\eqref{y_solution_CIS}, we obtain four linear algebraic equations for unknowns $\alpha_M$ that can be written in the matrix form as
\begin{equation}
C\alpha_M^\top=\omega\alpha_M^\top,
\end{equation} 
where the superscript $\top$ is the transpose operator, and the $4\times4$ matrix $C$ reads
\begin{equation}\label{CIS}
C=\left[ \begin {array}{cccccc} 
\int{y_0d\bm{v}} & 2\int{v_1y_0d\bm{v}}& 2\int{v_2y_0d\bm{v}} & \int{Vy_0d\bm{v}} 
\\ \noalign{\medskip}
\int{v_1y_0d\bm{v}} & 2\int{v_1^2y_0d\bm{v}}& 2\int{v_1v_2y_0d\bm{v}} & \int{v_1Vy_0d\bm{v}}
\\ \noalign{\medskip}
\int{v_2y_0d\bm{v}} & 2\int{v_1v_2y_0d\bm{v}}& 2\int{v_2^2y_0d\bm{v}} & \int{v_2Vy_0d\bm{v}}
\\ \noalign{\medskip}
\int{Vy_0d\bm{v}} & 2\int{Vv_1y_0d\bm{v}}& 2\int{Vv_2y_0d\bm{v}} & \int{V^2y_0d\bm{v}} 
\end {array} \right], 
\end{equation}
with $V=v^2-1$ and
\begin{equation}
y_0(\bm{v})=\frac{E(\bm{v})}{1+iK\bm{\theta}\cdot\bm{v}}.
\end{equation}

\begin{figure}[t]
	\centering
	\includegraphics[scale=0.45]{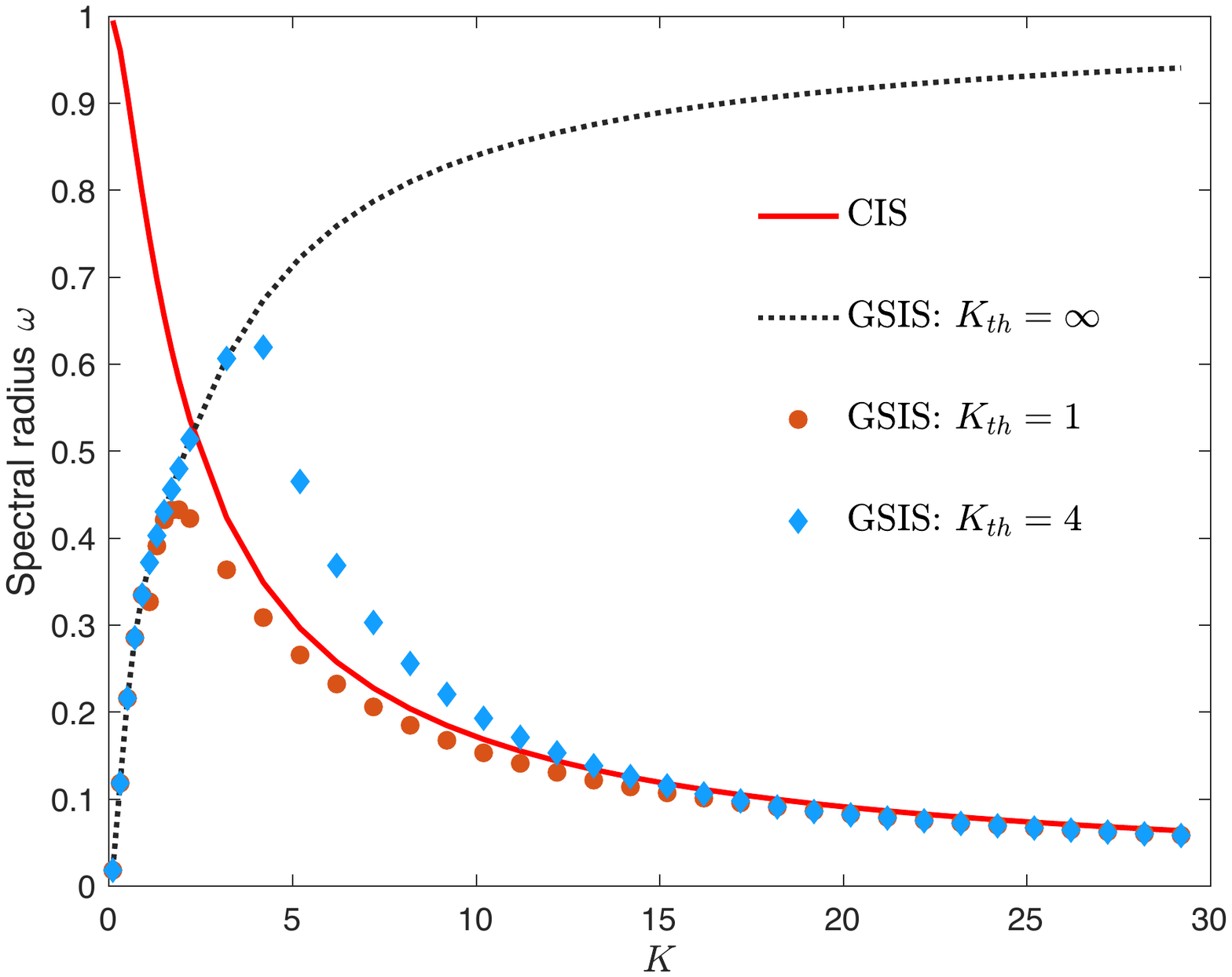}
	\caption{ The spectral radius $\omega$ as a function of the Knudsen number $K$ in both CIS and GSIS. }
	\label{fig:SR}
\end{figure}

In general, the spectral radius (i.e. the convergence rate, or the error decay rate between two consecutive iterations) can be obtained by numerically computing the eigenvalues $\omega$ of matrix $C$ and taking the maximum absolute value of $\omega$. The result of spectral radius as a function of the Knudsen number is shown in Figure~\ref{fig:SR}, from which we find that in the limiting case of $K\rightarrow0$, the spectral radius satisfies
\begin{equation}
\omega_{CIS}\rightarrow1-\frac{K^2}{2}.
\end{equation}

The fact that $\omega_{CIS}$ approaches one when $K\rightarrow0$ means that errors defined in Eqs.~\eqref{Y_expression} and~\eqref{Phi_expression} decay rather slowly. Worse still, the CIS encounters the problem of ``false convergence''. According to the analysis of Adam and Larsen for radiation transfer problem~\cite{DSA2002} (same for the BGK equation here), if the iteration is terminated at the $(k+1)$-th step with the following convergence criterion
\begin{equation}\label{false_convergence}
|\Phi_M^{(k+1)}-\Phi_M^{(k)}|<\epsilon,
\end{equation}
then the relative difference from the true steady-state solution $\Phi_M$ of Eq.~\eqref{LBE_BGK} satisfies
\begin{equation}\label{false_convergence2}
|\Phi_M^{(k+1)}-\Phi_M|<\frac{\omega_{CIS}}{1-\omega_{CIS}}\epsilon\rightarrow\frac{2\epsilon}{K^2}, \quad \text{when~}K\rightarrow0.
\end{equation}
The asymptotic expression at $K\rightarrow0$ shows that, if the convergence criterion is chosen as Eq.~\eqref{false_convergence}, the error in final iteration can be much larger than the preassigned value $\epsilon$; thus false convergence occurs if $\epsilon$ is not small enough. To reach the same convergence criterion for $|\Phi_M^{(k+1)}-\Phi_M|$, the total number of iterations must scale as $O(K^{-N})$ with $N>2$. This proves that in CIS it is very hard to obtain the converged solution in the near-continuum flow regime; one numerical example is given in Figure 3 of Ref.~\cite{WANG201833}.

\section{GSIS and its convergence rate}\label{GSIS:section}
In GSIS, additional macroscopic equations are coupled with CIS to boost convergence~\cite{SuArXiv2019}. The algorithm is summarized below. First, given the value of velocity distribution function $h^{(k)}(\bm{x},\bm{v})$ at the $k$-th iteration step, the velocity distribution function at the intermediate $(k+1/2)$-th step is obtained by solving the following kinetic equation: 
\begin{equation}\label{LBE_iteration_2}
h^{(k+1/2)}+
K\bm{v}\cdot\frac{\partial
	{h}^{(k+1/2)}}{\partial{\bm{x}}}=h_{eq}(h^{(k)}).
\end{equation}

Second, macroscopic quantities $\bar{M}=(\bar{\varrho},\bar{u}_1,\bar{u}_2, \bar{\tau})$ are introduced to update $M$ at the (k+1)-th iteration step in the following manner:
\begin{equation}\label{GSIS_K30}
M^{(k+1)}=\beta\bar{M}+(1-\beta)M^{(k+1/2)},
\end{equation} 
where $0\le\beta\le1$ is the relaxation coefficient: when $\beta=0$ the scheme is reduced to the CIS with no fast convergence at small values of Knudsen number. The macroscopic quantities $\bar{M}$ can be obtained by solving the following partial differential equations (note that the Einstein summation is used):
\begin{eqnarray}
\frac{\partial {\bar{u}_i}}{\partial{x_i}}=0, \label{macro_1} \\
\frac{\partial {\bar{\varrho}}}{\partial{x_i}}+\frac{\partial {\bar{\tau}}}{\partial{x_i}}+\frac{\partial {\bar{\sigma}_{ij}}}{\partial{x_j}}=0, \label{macro_2} \\
\frac{\partial {\bar{q}_i}}{\partial{x_i}}=0, \label{macro_3}
\end{eqnarray}
which are obtained by taking the velocity moments of the BGK equation~\eqref{LBE_BGK}. However, in GSIS the shear stress $\bar{\sigma}_{ij}$ and heat flux $\bar{q}_i$ are not directly calculated from the velocity distribution function $h^{(k+1/2)}$ according to Eqs.~\eqref{NS_shear_def} and~\eqref{NS_heat_def}, but are reconstructed as
\begin{eqnarray}
\bar{\sigma}_{ij} =-2K_e\frac{\partial \bar{u}_{<i}}{\partial {x_{j>}}}+\text{HoT}_{\sigma_{ij}}, \label{sigma_HoT}\\
\bar{q}_i =-K_e\frac{\partial \bar{\tau}}{\partial x_i}+\text{HoT}_{q_i}, \label{q_HoT}
\end{eqnarray} 
such that Eqs.~\eqref{macro_1}, \eqref{macro_2} and~\eqref{macro_3} are reduced to the linearized Navier-Stokes equations when the high-order terms $\text{HoT}_{\sigma_{ij}}$ and $\text{HoT}_{q_i}$ vanish, and when the effective Knudsen number $K_e$ takes the value of $K_e=K$; the role of $K_e$ will be discussed below in section~\ref{sec:whyimportant}, where we will show the inclusion of Newton's law for shear stress and Fourier's law for heat conduction in Eqs.~\eqref{sigma_HoT} and~\eqref{q_HoT} is critical to realize the fast convergence for near-continuum flows. Note that the high-order terms $\text{HoT}_{\sigma_{ij}}$ and $\text{HoT}_{q_i}$ should be derived exactly from the kinetic equation~\eqref{LBE_BGK}, so that the rarefaction effects beyond the Navier-Stokes level are all captured.

There are many ways to construct the expressions for these high-order terms. In our  paper~\cite{SuArXiv2019}, they are constructed by multiplying Eq.~\eqref{LBE_BGK} with $\left(v_iv_j-{v^2}\delta_{ij}/2\right)E$ and ${v_i}\left(v^2-2\right)E$, respectively, and integrating the resultant equations with respect to the molecular velocity $\bm{v}$: 
\begin{eqnarray}
\text{HoT}_{\sigma_{ij}}=2K_e\frac{\partial u^{(k+1/2)}_{<i}}{\partial {x_{j>}}}-2K\int{\left(v_iv_j-\frac{v^2}{2}\delta_{ij}\right)}\bm{v}\cdot\frac{\partial{} h^{(k+1/2)}}{\partial\bm{x}}d\bm{v}, \label{higher_order_app1}\\
\text{HoT}_{q_i} =K_e\frac{\partial \tau^{(k+1/2)}}{\partial x_i}-K\int{v_i(v^2-2)}\bm{v}\cdot\frac{\partial{} h^{(k+1/2)}}{\partial\bm{x}}d\bm{v}. \label{higher_order_app2}
\end{eqnarray}

Clearly, the above high-order terms are correct when the steady state is reached. However, during iteration when the steady state is not reached, Eqs.~\eqref{sigma_HoT} and~\eqref{q_HoT} are not correct since macroscopic quantities $\bar{M}$ are evaluated at the $(k+1)$-th iteration step, while high-order terms $\text{HoT}_{\sigma_{ij}}$ and $\text{HoT}_{q_i}$ are computed using the intermediate velocity distribution function at the $(k+1/2)$-th iteration step. 

\subsection{The case of $M^{(k+1)}=\bar{M}$ and $K_e=K$} To calculate the convergence rate of GSIS, we introduce the error function for the velocity distribution function as
\begin{eqnarray}
Y^{(k+1/2)}(\bm{x},\bm{v})=h^{(k+1/2)}(\bm{x},\bm{v})-h^{(k)}(\bm{x},\bm{v})=\omega^{k}y(\bm{v})\exp(i\bm{\theta}\cdot{\bm{x}}),\label{Y_ansatz2} 
\end{eqnarray}
while the error functions $\Phi_M$ are still defined as $\Phi^{(k+1)}_M(\bm{x})=M^{(k+1)}(\bm{x})-M^{(k)}(\bm{x})$. It is clear that, according to Eq.~\eqref{LBE_iteration_2}, the solution of $y(\bm{v})$ is still given by Eq.~\eqref{y_solution_CIS}. It should be emphasized that, in GSIS the error functions $\Phi_M$ are not directly calculated from $Y$. Instead, macroscopic quantities $M^{(k+1)}(\bm{x})$  are governed by Eqs.~\eqref{macro_1}, \eqref{macro_2} and~\eqref{macro_3}, from which one can obtain the governing equations for $\Phi^{(k+1)}_M(\bm{x})$.

On substituting Eqs.~\eqref{Y_ansatz2}, \eqref{Phi_expression} and  \eqref{Phi_ansatz} into the governing equations for $\Phi^{(k+1)}_M(\bm{x})$, we obtain the following four linear algebraic equations:
\begin{eqnarray}
\omega(i\theta_1\alpha_{u_1}+i\theta_2\alpha_{u_2})=0, \label{GSIS_first}\\
\omega[i\theta_1(\alpha_\varrho+\alpha_\tau)+K_e\alpha_{u_1}]=S_1,\\
\omega[i\theta_2(\alpha_\varrho+\alpha_\tau)+K_e\alpha_{u_2}]=S_2,\\
\omega{K_e}\alpha_{\tau}=S_\tau, \label{GSIS_middle1}
\end{eqnarray}
where the source terms, due to high-order terms in Eqs.~\eqref{higher_order_app1} and~\eqref{higher_order_app2}, are also linear functions of $\alpha_M$: $S_1=\int{s_1}yEd\bm{v}$, $S_2=\int{s_2}yEd\bm{v}$, and $S_\tau=\int{s_\tau}yEd\bm{v}$, with
\begin{eqnarray}
s_1=K_ev_1+K(i\theta_1v_1+i\theta_2v_2)\left[i\theta_1(v_1^2-v_2^2)+2i\theta_2v_1v_2\right], \label{GSIS_middle2}\\
s_2=K_ev_2+K(i\theta_1v_1+i\theta_2v_2)\left[2i\theta_1v_1v_2+i\theta_2(v_2^2-v_1^2)\right], \\
s_\tau=K_eV+K(i\theta_1v_1+i\theta_2v_2)\left(i\theta_1v_1+i\theta_2v_2\right)(V-1). \label{GSIS_end}
\end{eqnarray}

The spectral radius of GSIS can be obtained by solving Eqs.~\eqref{GSIS_first}-\eqref{GSIS_end}. That is, these equations are firstly rewritten in the matrix form as
\begin{equation}
L\omega\alpha_M^\top=R\alpha_M^\top,
\end{equation} 
where the $4\times4$ matrix $L$ is obtained from the left-hand side of equations~\eqref{GSIS_first}-\eqref{GSIS_middle1} as:
\begin{equation}\label{GSIS_L}
L=\left[ \begin {array}{cccccc} 
0& i\theta_1& i\theta_2 &0 
\\ \noalign{\medskip}
i\theta_1 &K_e &0 &i\theta_1 
\\ \noalign{\medskip}
i\theta_2 &0 &K_e &i\theta_2 
\\ \noalign{\medskip}
0  &0  & 0 &K_e
\end {array} \right], 
\end{equation}
and due to the fact that $y(\bm{v})$ in Eq.~\eqref{y_solution_CIS} a linear combination of $\alpha_\varrho, \alpha_{u_1}, \alpha_{u_2}$ and $\alpha_{\tau}$, the $4\times4$ matrix $R$ is obtained from Eqs.~\eqref{GSIS_middle2}-\eqref{GSIS_end} as 
\begin{equation}\label{GSIS_R}
R=\left[ \begin {array}{cccccc} 
0& 0 & 0 & 0 
\\ \noalign{\medskip}
\int{y_0s_1d\bm{v}} & 2\int{v_1y_0s_1d\bm{v}}& 2\int{v_2y_0s_1d\bm{v}} & \int{Vy_0s_1d\bm{v}}
\\ \noalign{\medskip}
\int{y_0s_2d\bm{v}} & 2\int{v_1y_0s_2d\bm{v}}& 2\int{v_2y_0s_2d\bm{v}} & \int{Vy_0s_2d\bm{v}}
\\ \noalign{\medskip}
\int{y_0s_\tau{}d\bm{v}} & 2\int{v_1y_0s_\tau{}d\bm{v}}& 2\int{v_2y_0s_\tau{}d\bm{v}} & \int{Vy_0s_\tau{}d\bm{v}}
\end {array} \right].
\end{equation}

By introducing $G=L^{-1}R$ and numerically computing the eigenvalues of the matrix $G$ we obtain the spectral radius $\omega$ of GSIS. The results are shown in the dotted black line in Figure~\ref{fig:SR}. It is seen that the spectral radius goes to zero as $\omega\sim{K^2}$ when $K\rightarrow0$. This demonstrates that the GSIS is able to boost convergence significantly in the near-continuum flow regime. However, the spectral radius increases to one when $K\rightarrow\infty$. This stands in sharp contrast to DSA schemes for rarefied Poiseuille-type flows, where the flow velocity (only in one direction) is perpendicular to spatial variables~\cite{Valougeorgis2003}. As a consequence, the synthetic equation for the only flow velocity is of diffusion-type~\cite{Valougeorgis2003,LeiJCP2017,SU2019573}, so that the spectral radius goes to zero when $K\rightarrow\infty$ and $K\rightarrow0$. However, in GSIS we have multiple flow velocities whose directions are inside the spatial domain, and the synthetic equations contain the continuity equation~\eqref{macro_1} which is not of diffusion type. We believe the continuity equation makes the spectral radius goes to one when the Knudsen number is large.

To make the spectral radius in GSIS goes to zero when Knudsen number is large, in Eq.~\eqref{GSIS_K30} we choose the relaxation coefficient as
\begin{equation}\label{GSIS_K3}
\beta=\frac{min(K,K_{th})}{K},
\end{equation}
where $K_{th}$ is the threshold Knudsen number, so that GSIS is reduced to CIS when $K$ is large, where the spectral radius approaches zero. The spectral radius of this GSIS can be obtained by computing the eigenvalue of the matrix $G=\beta{L^{-1}R}+(1-\beta)C$, 
where the results at the threshold Knudsen number of values 1 and 4 are shown in Figure~\ref{fig:SR}. Clearly, by choosing approximate value of $\beta$, we can make the maximum value of $\omega$ less than 0.5; this means that after 10 iterations, the error will be decreased by at least three orders of magnitude. Thus, theoretically, GSIS can reach fast convergence in the whole range of Knudsen number, which has been demonstrated by the numerical examples in our recent paper~\cite{SuArXiv2019}.

In addition to the fast converge at large Knudsen numbers, Eq.~\eqref{GSIS_K3} with $\beta<1$ can make the algorithm more stable. This is because the high-orders terms defined in Eqs.~\eqref{higher_order_app1} and~\eqref{higher_order_app2} may have strong variations around sharp (i.e. rectangular) corners, which leads to unphysical variations of density, velocity and temperature when solving Eqs.~\eqref{macro_1}, \eqref{macro_2} and~\eqref{macro_3}. By updating the macroscopic quantities according to Eq.~\eqref{GSIS_K30}, the error at sharp boundaries is reduced and stable algorithm is guaranteed~\cite{SuArXiv2019}.

\begin{figure}[t]
	\centering
	\includegraphics[scale=0.45]{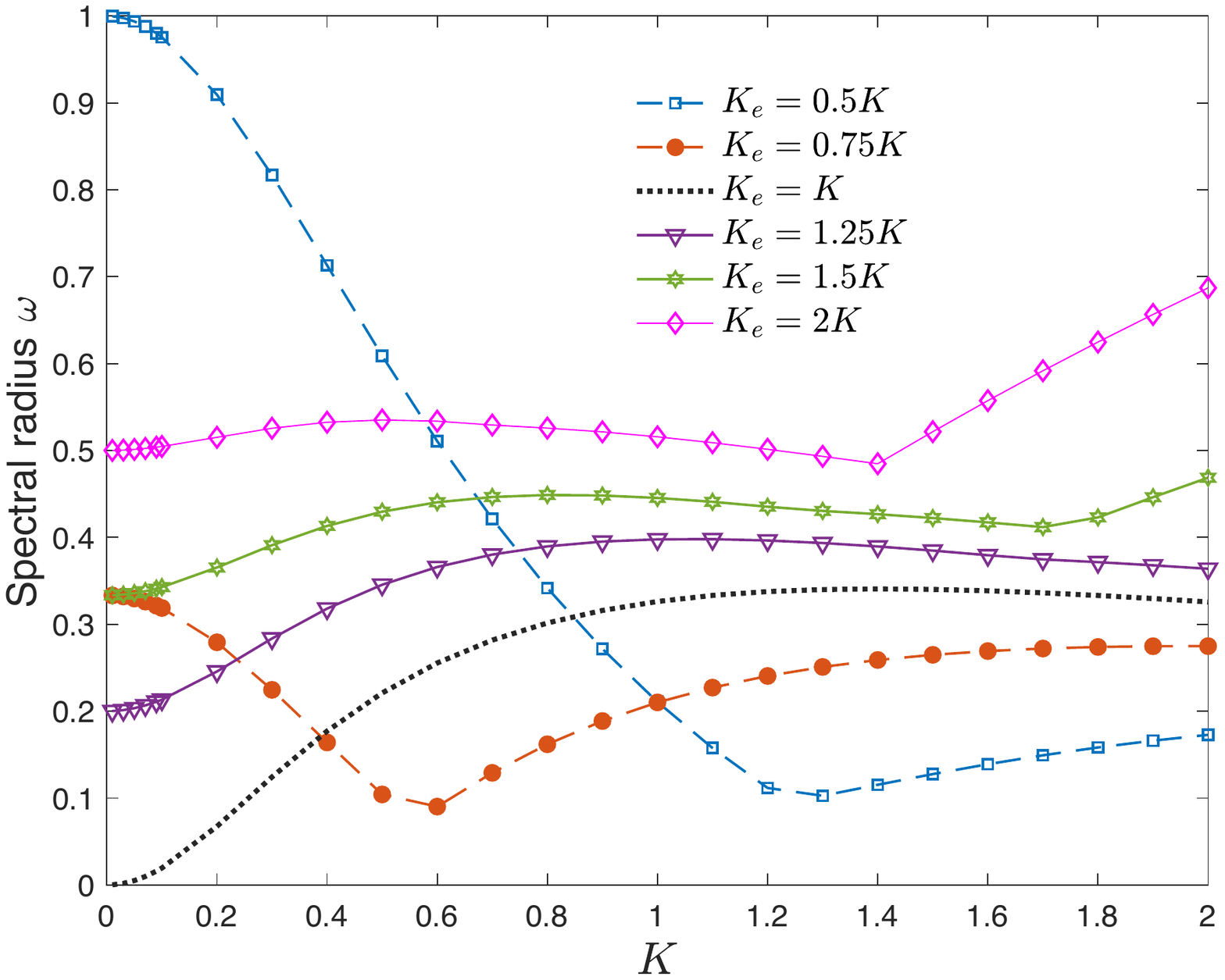}
	\caption{ The spectral radius $\omega$ as a function of the Knudsen number $K$ in GSIS, when $K_e$ in Eqs.~\eqref{sigma_HoT} and~\eqref{q_HoT} takes different values and the relaxation coefficient  in Eq.~\eqref{GSIS_K30} is $\beta=1$. }
	\label{fig:whyimportant}
\end{figure}

\subsection{Why are the Navier-Stokes constitutive relations important}\label{sec:whyimportant}
Note that when the Knudsen number is small, we choose $K_e=K$ in Eqs.~\eqref{sigma_HoT} and~\eqref{q_HoT}; this means that in the near-continuum regime, the constitutive relations in Navier-Stokes equations, i.e. Newton's law for shear stress and Fourier's law for heat conduction are explicitly included in the macroscopic synthetic equations. This turns out to be extremely important for the fast convergence of GSIS, as the spectral radius goes to zero when $K\rightarrow0$. To demonstrate this superiority, we calculate the spectral radius by choosing different values of $K_e$ when the Knudsen number is small; the results in Figure~\ref{fig:whyimportant} show that only when $K_e=K$ can the spectral radius go to zero when $K\rightarrow0$. Moreover, when the Knudsen is small, e.g. $K<0.2$, the further the effective Knudsen number $K_e$ deviates away from $K$, the larger the spectral radius and hence the slower the convergence. These results confirm that the fastest convergence can be realized by including the Navier-Stokes constitutive relations explicitly in synthetic equations. If the shear stress and heat flux in Eqs.~\eqref{macro_2} and~\eqref{macro_3} are directly computed from the $h^{(k+1/2)}$ according to Eqs.~\eqref{NS_shear_def} and~\eqref{NS_heat_def}, our Fourier stability analysis finds that the scheme does not permit fast convergence in the near-continuum flow regime.


\section{Asymptotic preserving property of GSIS}\label{AP_property} Note that in GSIS the kinetic equation and macroscopic synthetic equations can be solved by different numerical methods with different order of accuracy. Now we consider the influence of spatial discretization in the gas kinetic solver on the accuracy of GSIS, based on the assumptions that (i) synthetic equations can be solved exactly and (ii) the spatial cell size $\Delta{x}$ is able to capture the physical solution. To this end, we consider whether the Navier-Stokes equations can be exactly derived, through the Chapman-Enskog expansion~\cite{CE}, from the discretized gas kinetic equation with the following scaling:
\begin{equation}\label{scaling}
\Delta{x}\sim{K^{1/\alpha}}.
\end{equation}
Here $\alpha$ denotes the order of accuracy in the asymptotic preserving of Navier-Stokes equations. Clearly, the larger the value of $\alpha$, the better the numerical scheme, as we can use less number of spatial cells to achieve the same accuracy. It has been rigorously proven that the discrete unified gas kinetic scheme has $\alpha=2$, so that $\Delta{x}$ can be much larger than the mean free path to capture the hydrodynamics in bulk regions~\cite{Guo2019UP_arXiv}. If $\alpha=\infty$, the scheme will capture the hydrodynamical behavior of gas when $\Delta{x}$ is approximately the system size (no matter what the value of $K$ is), as long as this size is adequate to capture the flow physics, for example the dynamics of a sound wave whose wavelength is approximately the system size.

Since the spectral radius of GSIS approaches zero when $K\rightarrow0$, the converged solution can be found within a few number of iterations, and we have $h_{eq}^{(k+1)}=h_{eq}^{(k)}$ and $M^{(k+1)}=M^{(k)}$. Thus, when the iteration is converged, the kinetic equation~\eqref{LBE_iteration_2} in the discretized form can be written as
\begin{equation}\label{LBE_GSIS}
\bm{v}\cdot\frac{\partial{h}}{\partial\bm{x}}+O(\Delta{x}^n)\delta(h)=\frac{h_{eq}-h}{K},
\end{equation}
where  $n$ is the order of approximation for the spatial derivative $\partial {h}/\partial{x}$, and $\delta(h)$ is the $(n+1)$-th order derivative of $h$. For instance, if we use the second-order upwind finite difference scheme, $n=2$. If the third-order approximating polynomials are used in the DG method~\cite{WeiSuJCP1,Su2019IDG}, we have $n=4$. 


In the Chapman-Enskog expansion the velocity distribution function is approximated by the Taylor expansion
$h=h_0+Kh_1+\cdots$.
By subsisting this expansion into Eq.~\eqref{LBE_GSIS} and collecting terms with the order of $K^{-1}$, we have $h_0=f_{eq}$, when the following largest scaling is chosen
\begin{equation}\label{scaling2}
\Delta{x}\sim{K^{1/\infty}}=O(1).
\end{equation}
Under the scaling~\eqref{scaling2}, by collecting terms with the order $K^{0}$, we have
$h_1=-\bm{v}\cdot{\partial{h_{eq}}}/{\partial\bm{x}}-\delta(h_{eq})$.
Thus, according to Eqs.~\eqref{sigma_HoT}, \eqref{q_HoT}, \eqref{higher_order_app1} and~\eqref{higher_order_app2}, Newton's law for shear stress and Fourier's law for heat conduction are recovered in macroscopic synthetic equations with accuracy $O(K^2)$:
\begin{eqnarray}
\sigma_{ij} =-2K\frac{\partial u_{<i}}{\partial {x_{j>}}}+O(K^2), \quad
q_i =-K\frac{\partial \tau}{\partial x_i}+O(K^2).
\end{eqnarray} 

Thus, as long as the spatial resolution $\Delta{x}=O(1)$ is able to capture the physical solution, GSIS is able to recover the linearized Navier-Stokes equations when $K\rightarrow0$. In this sense the scheme is better than the discrete unified gas kinetic scheme where $\Delta{x}=O(\sqrt{K})$~\cite{Guo2019UP_arXiv}. Therefore, the overall order of accuracy of GSIS depends only on the order to solve macroscopic synthetic equations, that is, $n$ in Eq.~\eqref{LBE_GSIS}. In reality, however, such a large spatial cell size $\Delta{x}=O(1)$ cannot be used in regions with Knudsen layers or shock structures, where the physical solutions require a spatial resolution of $O(K)$. Fortunately, these kinetic layers only take up a small fraction of the computational domain, say, in the vicinity of solid walls, which can be captured by implicit schemes with non-uniform spatial discretization. This will be tested in the following numerical examples.

\section{Numerical examples}\label{sec_num}

To assess the analytical results in previous sections, we simulate the one-dimensional coherent Rayleigh-Brillouin scattering, sound wave propagation, and Couette flows, as well as two-dimensional thermally-induced flow. 

\subsection{Coherent Rayleigh-Brillouin scattering} In coherent Rayleigh-Brillouin scattering the wavelike density perturbation in gas is created by a moving optical lattice, where the governing equation for the velocity distribution function $h'$ can be written as 
\begin{equation}\label{CRBS_LBE}
\frac{\partial {h'}}{\partial {t}}+v_2\frac{\partial{h'}}{\partial{x_2}}=\frac{h'_{eq}-h'}{K}+2v_2\cos(2\pi{x_2}+2\pi{f_s}t),
\end{equation}
where $f_s$ is the frequency of the moving optical lattice; we choose $f_s=\sqrt{5/6}$ (i.e. the sound speed normalized by the most probable speed of gas molecules) so that the amplitude of perturbed density will scale as $1/K$ when the Knudsen number is small. If the numerical scheme cannot preserve the  Navier-Stokes limit when $K\rightarrow0$, then a huge number of spatial cells is needed to keep a low numerical dissipation.  Since the interaction between the gas and solid wall is absent~\cite{WuLei_RG}, this problem is ideal to test the accuracy and efficiency of the kinetic scheme.

The problem is periodic in both temporal and spatial directions. Therefore, using the transform $h'(x_2,\bm{v},t)=h(x_2,\bm{v})\exp(2\pi{}i{f_s}t)$, the final perturbed density profile can be obtained by solving the following equation~\cite{WuLei_RG}:
\begin{equation}\label{CRBS_govering}
2\pi{i}f_sh+v_2\frac{\partial{h}}{\partial{x_2}}=\frac{h_{eq}-h}{K}+2v_2\cos(2\pi{x_2}),
\end{equation}
which can be handled by implicit solvers.

Note that in order to keep the calculation more realistic, in the following the linearized Shakhov kinetic model is used and the velocity distribution function contains the three-dimensional molecular velocity space~\cite{Shakhov1968}. That is, the reference distribution function in Eq.~\eqref{CRBS_govering} now becomes
\begin{equation}\label{LBE_BGK_gain3}
h_{eq}(\bm{x},\bm{v})=\varrho(\bm{x})+2\bm{u}(\bm{x})\cdot\bm{v}+\tau(\bm{x})\left(v^2-1\right)+\frac{4}{15}\bm{q}\cdot\bm{v}\left(v^2-\frac{5}{2}\right),
\end{equation}
where the perturbed density is $\rho=\int {}E_3(\bm{v})hd\bm{v}$ with the equilibrium velocity distribution function ${E}_3(\bm{v})=\exp(-v^2)/\pi^{1.5}$, the flow velocity is $\bm{u}=\int \bm{v}E_3(\bm{v})hd\bm{v}$, the perturbed temperature is $\tau=\int (2v^2/3-1)E_3(\bm{v})hd\bm{v}$, and the heat flux is $\bm{q}=\int \bm{v}\left(v^2-5/2\right)E_3(\bm{v})hd\bm{v}$.
	
The presence of the first term in Eq.~\eqref{CRBS_govering} modifies the
synthetic equations~\eqref{macro_1}, \eqref{macro_2} and~\eqref{macro_3} to the following forms~\cite{SuArXiv2019}: 
\begin{eqnarray}
2\pi{i}f_s\bar{\varrho}+\frac{\partial {\bar{u}_i}}{\partial{x_i}}=0, \\
2\pi{i}f_s\bar{u}_i+\frac{\partial {\bar{\varrho}}}{\partial{x_i}}+\frac{\partial {\bar{\tau}}}{\partial{x_i}}+\frac{\partial {\bar{\sigma}_{ij}}}{\partial{x_j}}=0, \\
2\pi{i}f_s\bar{\tau}+\frac{\partial {\bar{q}_i}}{\partial{x_i}}+\frac{\partial {\bar{u}_i}}{\partial{x_i}}=0, 
\end{eqnarray}
while the shear stress and heat flux are modified into the following forms:
\begin{eqnarray}
2\pi{i}f_s\sigma_{22}+\text{HoT}_{\sigma_{22}}+\frac{4}{3}\frac{\partial \bar{u}_2}{\partial {x_2}}=-\delta_\text{rp}\sigma_{22}, \label{sound_sigma} \\
2\pi{i}f_sq_{2}+\text{HoT}_{q_2}+\frac{5}{4}\frac{\partial \bar{\tau}}{\partial {x_2}}=-\frac{2}{3}\delta_\text{rp}q_{2}. \label{sound_q}
\end{eqnarray}

\begin{figure}[t]
	\centering
	\includegraphics[scale=0.3]{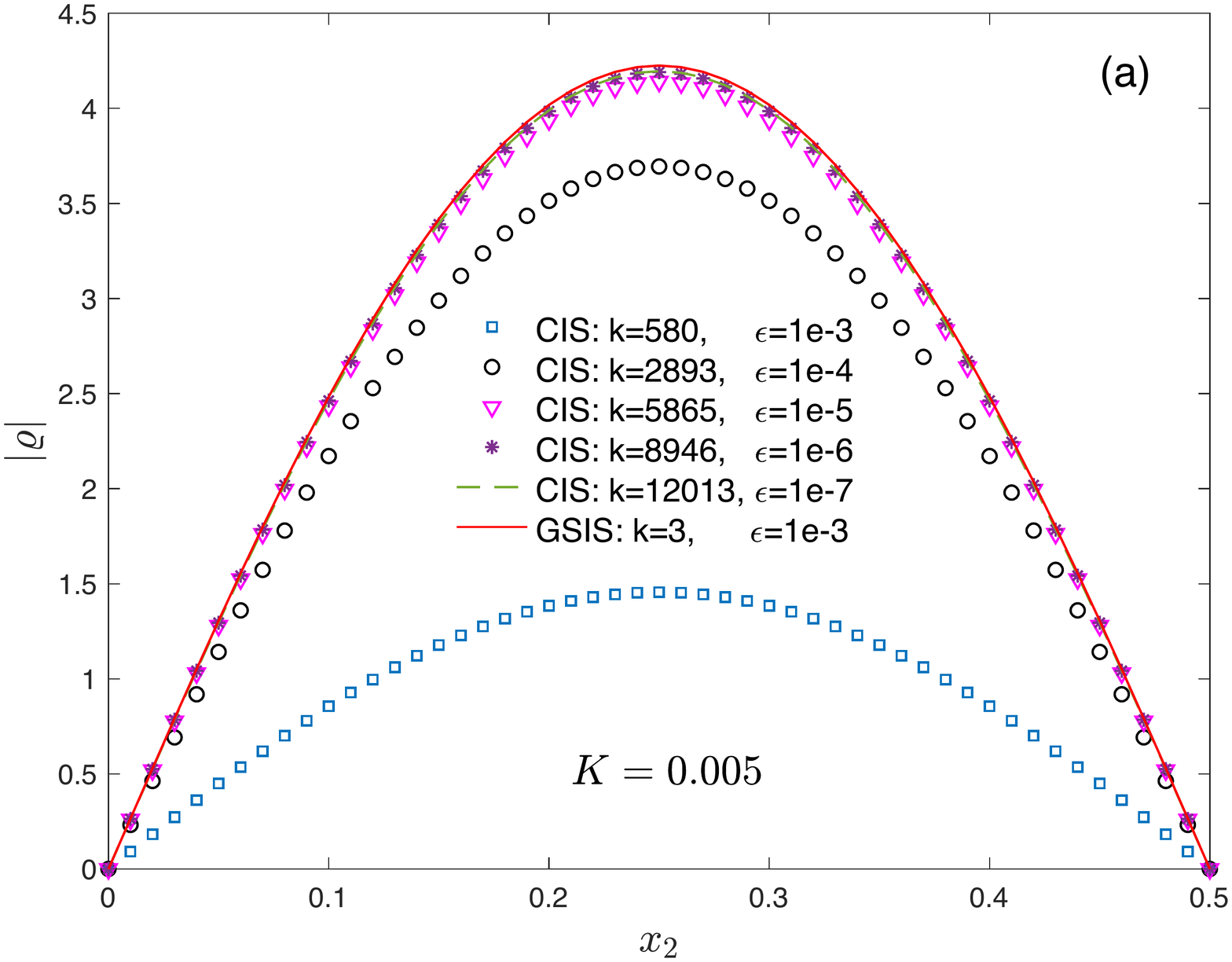}
	\includegraphics[scale=0.3]{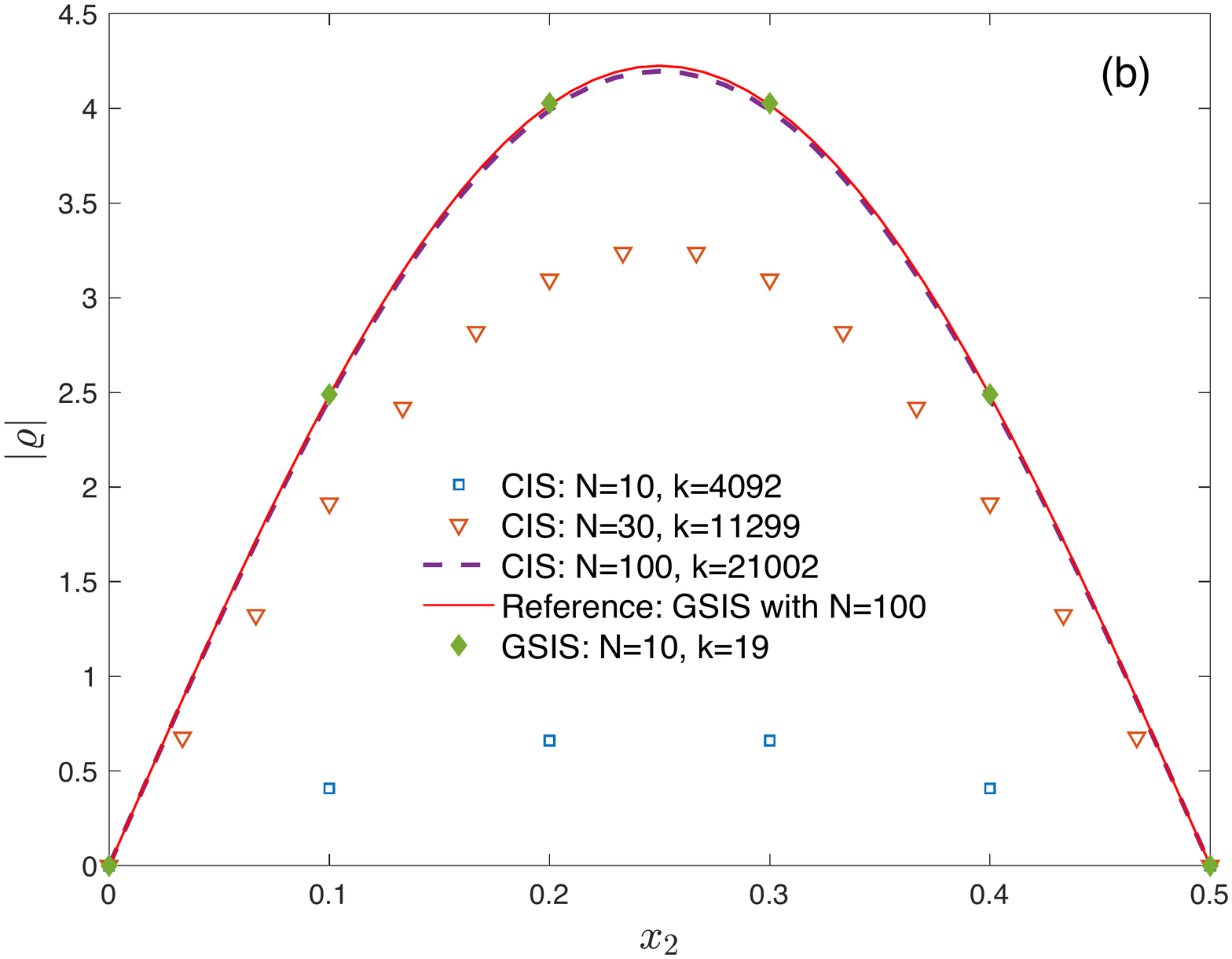}
	\caption{Comparisons of the density profile in coherent Rayleigh-Brillouin scattering obtained from CIS and GSIS, when the Knudsen number is $K=0.005$. (a) The convergence history of CIS, at different values of the error $\epsilon=\left|\int|\varrho^{(k+1)}dx_2|/\int|\varrho^{(k)}|dx_2-1\right|$. The spatial region $x_2\in[0,1]$ is divided into $N=100$ uniform cells, and due to symmetry only half of the density wave profile is plotted. The result from GSIS is converged even when $\epsilon=10^{-3}$, after three iterations. (b) The influence of spatial cell number on the density profile, where the iteration is terminated when $\epsilon<10^{-10}$. The kinetic equation is solved by the second-order upwind finite difference, while the synthetic equations are solved by the Fourier spectral method.  }
	\label{fig:CRBS_K_005}
\end{figure}

\begin{figure}[t]
	\centering
	\includegraphics[scale=0.45]{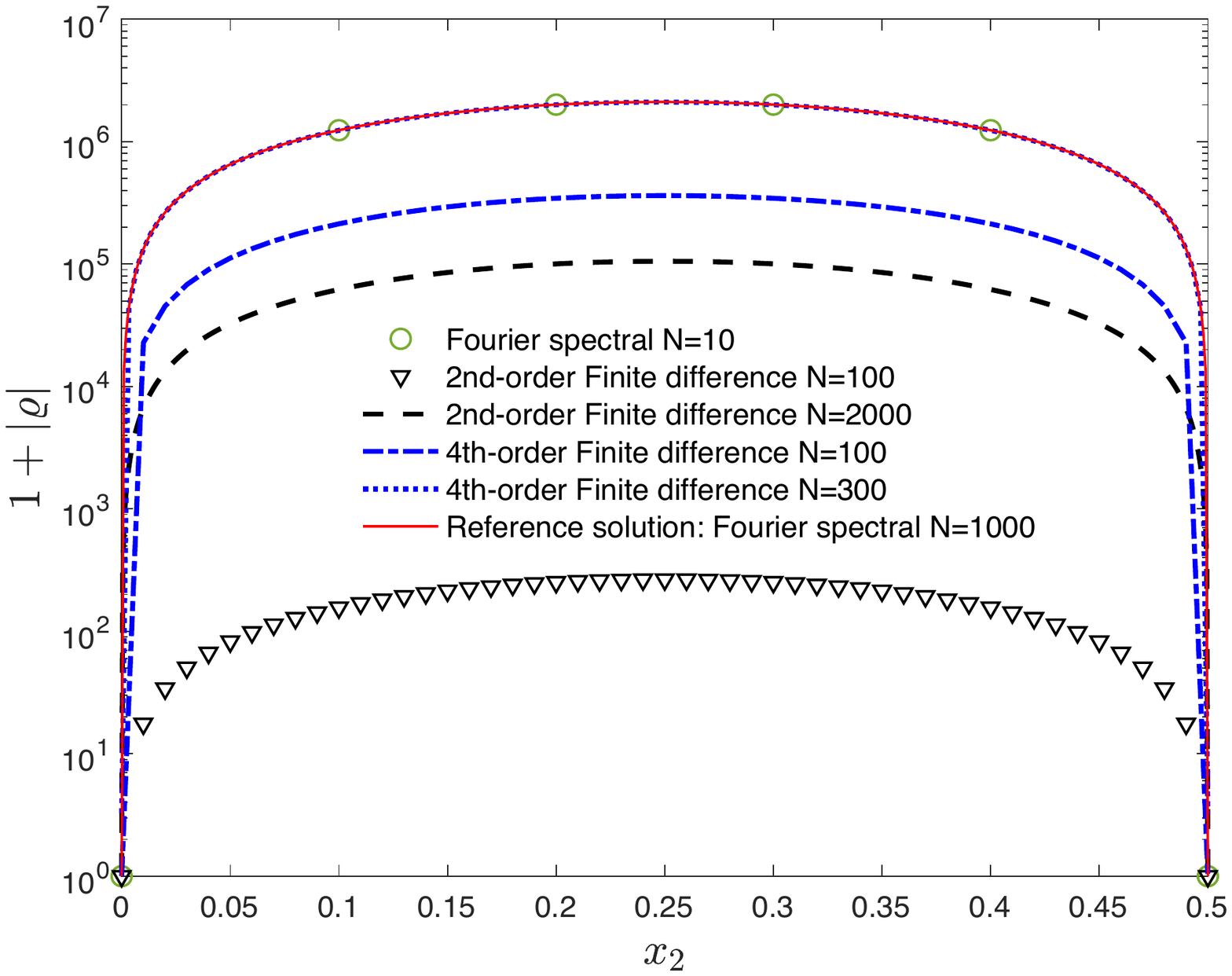}
	\caption{Comparisons of the density profile in coherent Rayleigh-Brillouin scattering, where the kinetic equation is solved by the second-order upwind finite difference, while macroscopic equations are solved by various schemes with different number of spatial points ($N$). The Knudsen number is $K=10^{-8}$. The density amplitude is shifted by one in order to show it in the log scale. }
	\label{fig:crbsKn1e-8}
\end{figure}

The comparison between CIS and GSIS in terms of efficiency and accuracy is shown in Figure~\ref{fig:CRBS_K_005}, when the Knudsen number is $K=0.005$ and the synthetic equations in GSIS are solved by the Fourier spectral method exactly. Due to the small value of convergence rate $\omega$, GSIS yields converged solution even after only 3 iterations and when $\epsilon=10^{-3}$, which is consistent with the analytical solution~\eqref{false_convergence2}. That is, 
\begin{equation}
|\Phi_M^{(k+1)}-\Phi_M|<\frac{\omega_{GSIS}}{1-\omega_{GSIS}}\epsilon\rightarrow\epsilon{K^2}, \quad \text{when~}K\rightarrow0.
\end{equation}
However, since $\omega_{CIS}$ approaches one when $K\rightarrow0$, CIS needs huge number of iterations and the convergence criterion has to be set very small, i.e. $\epsilon=10^{-10}$ in this case. This is also consistent with the prediction in Eq.~\eqref{false_convergence2}. In terms of spatial accuracy, Figure~\ref{fig:CRBS_K_005}(b) proves that GSIS can asymptotically preserve the Navier-Stokes limit with $\Delta{x}\sim{O(1)}$ provide this spatial resolution is adequate to describe the physical problem (here it is the density wave profile), while the CIS needs much smaller spatial cell size in order to keep the numerical dissipation small: when the cell size is large, the numerical dissipation leads to smaller amplitude of density wave when compared to the true solution.

To further show the importance of spatial resolution, we consider the same problem when the Knudsen number is $K=10^{-8}$ in Figure~\ref{fig:crbsKn1e-8}. Even in this case the Euler equation cannot be used, otherwise the amplitude of perturbation will go to infinity. The extremely small value of $K$ poses a real challenge to the kinetic scheme, as any small value of numerical dissipation could easily contaminate the final solution, which leads to a significant small amplitude of density perturbation. Even at such a small Knudsen number, if the macroscopic equation is solved exactly (by the Fourier spectral method) and $\Delta{x}\sim{O(1)}$ is able to capture the spatial variation, GSIS has infinite order of accuracy, see the circles in Figure~\ref{fig:crbsKn1e-8}. However, when the second-order finite difference scheme is used to solve the synthetic equations, we need more than 2000 spatial points to capture the density perturbation; this is understandable since the numerical dissipation (or numerical viscosity) $(\Delta{x})^2=2.5\times10^{-7}$ is larger than the physical viscosity (here it is reflected by the Knudsen number $K$). When the synthetic equations are solved by the fourth-order finite difference, we see that $N=100$ leads to wrong solutions, while $N=300$ yields accurate solution. This is because the numerical dissipations are about $(\Delta{x})^4=10^{-8}$ and $1.2\times10^{-10}$, respectively, so that the former is too dissipative while the latter is accurate.

\subsection{One-dimensional sound wave propagation}\label{sec:sound} 
\begin{figure}
	\centering
	\includegraphics[scale=0.31]{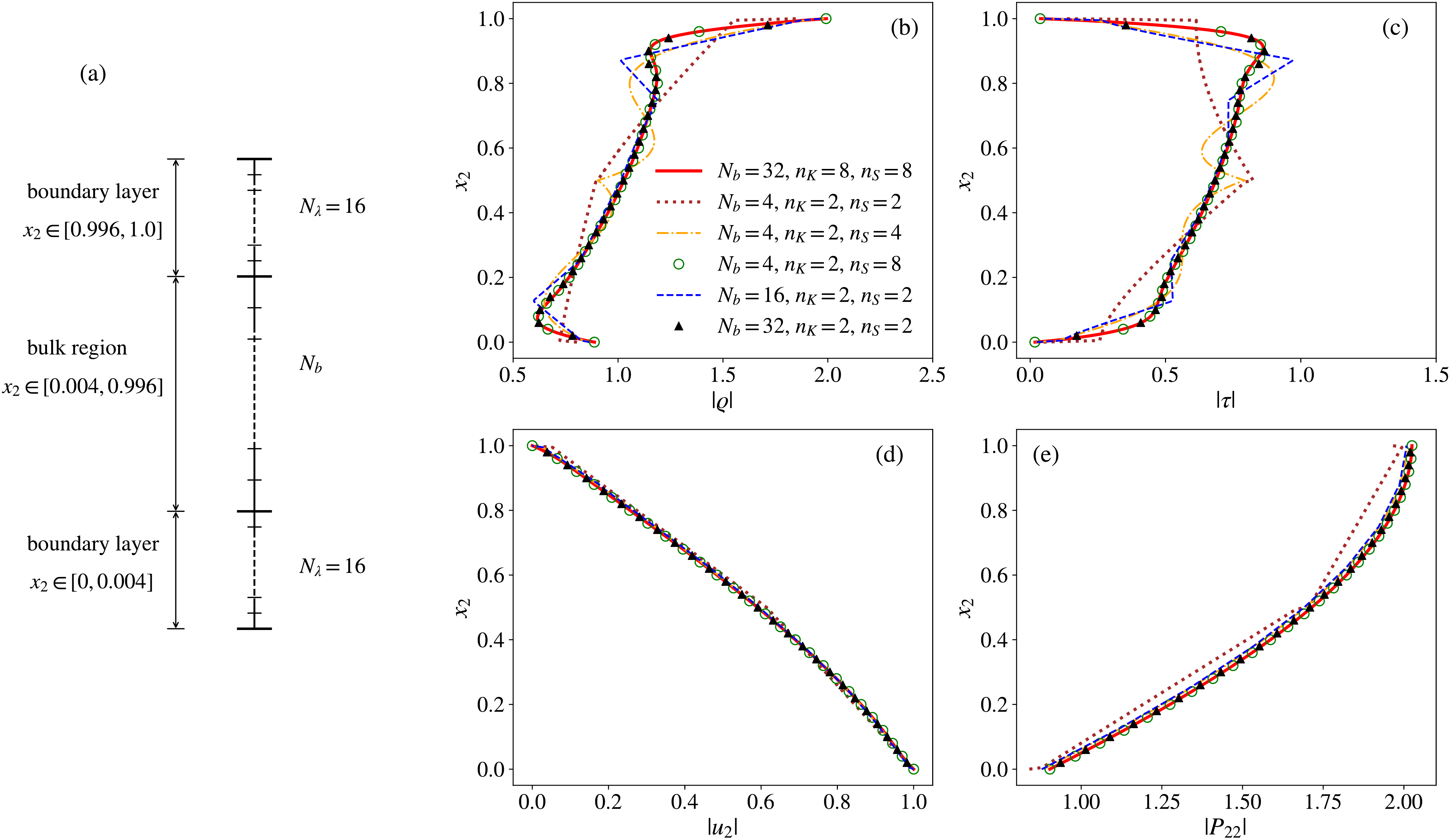}
	\caption{One-dimensional sound wave propagation induced by an oscillating wall at $x_2=0$, when the Knudsen number is $K=0.001$. (a) schematic illustration of the bulk region and boundary layer, and spatial discretization along $x_2$. (b)-(e) are respectively the amplitudes of density $|\varrho|$, temperature $|\tau|$, longitudinal velocity $|u_2|$ and normal pressure $|P_{22}|=|\sigma_{22}+\varrho+\tau|$ obtained from schemes with different orders of accuracy and numbers of cells. }
	\label{fig:SoundPropagation}
\end{figure}

The governing equation in this problem is the same as that in coherent Rayleigh-Brillouin scattering, except here we have no external driving force and the diffuse boundary condition is imposed on two walls: the one at $x_2=0$ is oscillating in the $x_2$ direction with a velocity $u_2=\cos(2\pi f_st)$ while the one at $x_2=1$ is stationary, see the schematic in Figure~\ref{fig:SoundPropagation}; detailed form of the diffuse boundary condition for the linearized kinetic equation can be found in Ref.~\cite{SuArXiv2019}. Here we choose $f_s=1/2\pi$ and $\sqrt{\pi}K/2=0.001$, and the kinetic and macroscopic equations are solved by DG with different orders of accuracy. This test is used to show that, when the Knudsen layer near the solid walls are well resolved, the accuracy of GSIS is mainly determined by the accuracy in solving macroscopic synthetic equations.  

In the numerical simulation we assume the thickness of Knudsen layer is about 4 times the mean free path in the immediate vicinity of walls~\cite{SU2019573}, which is divided into $N_{\lambda}=16$ equal cells each, see Figure~\ref{fig:SoundPropagation}(a). The bulk region $0.004<x_2<0.996$ is partitioned by $N_\text{b}$ uniform cells. The order of DG scheme for the kinetic equation is denoted by $n_\text{K}$ while the one for synthetic equations is $n_\text{S}$. In the constitutive relations~\eqref{sigma_HoT} and~\eqref{q_HoT}, the effective Knudsen number is $K_e=K$. During iteration, the macroscopic quantities $M^{(k+1)}$ are updated based on Eq.~\eqref{GSIS_K30}, and the relaxation coefficient is chosen as
\begin{equation}\label{Kn_loc}
\beta= \frac{min\left(K_\text{loc},K_{th}\right)}{K_\text{loc}},
\end{equation}
where $K_{th}$ is set as $min\left(1,\frac{5}{n_\text{K}-1}\right)$ and $K_\text{loc}$ is the local Knudsen number estimated from the local cell size: $K_\text{loc}=K/L_\text{loc}$ with $L_\text{loc}$ the minimum height of the triangles used in DG, where the one-dimensional problem is actually simulated on two-dimensional domain partitioned by structured triangles and the periodic boundary condition is imposed on the lateral boundaries of the domain. Note that the Knudsen number is defined as the mean free path over the characteristic system length $L$ and the spatial coordinate has been normalized by $L$. 

Figure~\ref{fig:SoundPropagation}(b)-(e) show the  amplitudes of density, temperature, velocity and normal pressure obtained from schemes with different orders of accuracy and numbers of cells. By choosing the numerical results with $N_\text{b}=32$ and $n_\text{K}=n_\text{S}=8$ as the reference, we see that the increase of $n_\text{S}$ or $N_b$ leads to solutions converge toward the reference one, even when the kinetic equation is solved with second-order accuracy, i.e. $n_\text{K}=2$. This means that in GSIS the accuracy is controlled by the macroscopic solver, at least in the near-continuum flow regime.

Table~\ref{Tab:SoundPropagation} shows the total number of iterations and the amplitude of normal pressure at the oscillating wall for different combinations of $N_\text{b}$, $n_\text{K}$ and $n_\text{S}$, when $\epsilon$ defined in Eq.~\eqref{general_epsilon} is less than $10^{-5}$. For all test cases, the convergence is reached within 50 iteration steps, which demonstrates the efficiency of GSIS, even when the streaming operator is discretized. The fact that the total iteration step is larger than that in the coherent Rayleigh-Brillouin scattering is that, in this case, although $K$ is small, the presence of Knudsen layer makes $K|\bm\theta|$ approximately one  in Eq.~\eqref{y_solution_CIS}. Thus, relative large iteration number is needed. On the other hand, Table~\ref{Tab:SoundPropagation} further confirms that the accuracy of solution in the bulk region is mainly determined by the accuracy in solving the macroscopic equations. As a consequence, the higher the order of DG for macroscopic equations, the more accurate the solution will be. Furthermore, accurate results can be obtained once  macroscopic quantities are well resolved on coarse mesh, where the cell size is much larger than the mean free path of gas molecules. For instance, when the 8th-order DG for macroscopic equations is employed with $N_b=4$, converged results are obtained on a mesh in the bulk region 248 times larger than the molecular mean free path. This is consistent with the results in section~\ref{AP_property} that  GSIS is able to asymptotically preserve the Navier-Stokes limit with a spatial size $\Delta{x}\sim{O}(1)$, provided this size is able to describe the physical solution.

\begin{table}
\caption{One-dimensional sound wave propagation when $\sqrt{\pi}K/2=0.001$: the number of iterative steps (Itr) when the residual $\epsilon<10^{-5}$ and the amplitude of normal pressure $|P_{22}|$ at the oscillating wall ($x_2=0$) under different combinations of $N_\text{b}$, $n_\text{K}$ and $n_\text{S}$. $N_\text{b}$ is the number of cells in the bulk region and $L_\text{b}=0.992/N_\text{b}$ is the cell size in the bulk region. $n_\text{K}$ and $n_\text{S}$ are the orders of DG to approximate the kinetic equation and synthetic equations, respectively.}

\centering
\begin{tabular}{ccccccccccc}
\hline
\multirow{3}{*}{$N_\text{b}$} & \multirow{3}{*}{$L_\text{b}$} & \multicolumn{6}{c}{$n_\text{K}=2$} & & \multicolumn{2}{c}{\multirow{2}{*}{$n_\text{K}=8$, $n_\text{S}=8$}} \\
\cline{3-9}
 & & \multicolumn{2}{c}{$n_\text{S}=2$} & \multicolumn{2}{c}{$n_\text{S}=4$} &  \multicolumn{2}{c}{$n_\text{S}=8$} & & \\
\cline{3-11}
&  & Itr & $|P_{22}|$ & Itr & $|P_{22}|$  & Itr & $|P_{22}|$ & & Itr & $|P_{22}|$ \\
\hline
4 & 0.248 & 46 & 0.840 & 36 & 0.882 &  32 & 0.903 & & 33 & 0.903 \\
8 & 0.124 & 47 & 0.846 & 36 & 0.900 &  32 & 0.903 & & 33 & 0.903 \\
16 & 0.062 & 47 & 0.877 & 36 & 0.903 & 32 & 0.903 & & 33 & 0.903 \\
32 & 0.031 & 47 & 0.897 & 36 & 0.903 & 32 & 0.903 & & 33 & 0.903 \\
\hline

\end{tabular}
\label{Tab:SoundPropagation}
\end{table}

\subsection{One-dimensional Couette flow} We consider the steady Couette flow between two infinite parallel plates located at $x_2=0$ and $x_2=1$. The top plate moves along the horizontal direction with a velocity of $u_1=1$, while the bottom plate moves in the opposite direction with the same magnitude of speed. The Knudsen number is set as $\sqrt{\pi}K/2=0.01$. The diffuse boundary condition is imposed on the two plates. This test is used to show the importance of resolving the Knudsen layer~\cite{SU2019573}.  

The partition of the computational domain along the longitudinal direction ($x_2$) is similar as that in Figure~\ref{fig:SoundPropagation}(a). In order to guarantee the accuracy of resolution in the bulk region, the bulk region $x_2\in[0.04,0.96]$ is partitioned into $N_\text{b}=32$ uniform cells, and the synthetic equations are solved with $N_\text{S}=8$.  Details to evaluate the constitutive relations in the synthetic equations and update the macroscopic quantities at $\left(k+1\right)$-th iteration step are given in section~\ref{sec:sound}.
We are interested in the velocity gradient at the center of the domain
\begin{equation}\label{velocity_gradient}
k_1=\frac{\mathrm{d}u_1(x_2=0.5)}{\mathrm{d}x_2},
\end{equation} 
as well as the Knudsen layer function $u_s$~\cite{SU2019573}:
\begin{equation}
u_s\left(x_2\right)=\frac{u_\text{NS}\left(x_2\right)-u_1\left(x_2\right)}{\sqrt{\pi}k_1K/2}
\label{KLF}
\end{equation}
where $u_\text{NS}$ is the velocity in the bulk region approximated by $u_\text{NS}\left(x_2
\right)=k_1\left(x_2-0.5\right)$.

\begin{figure}
	\centering
	\includegraphics[scale=0.4]{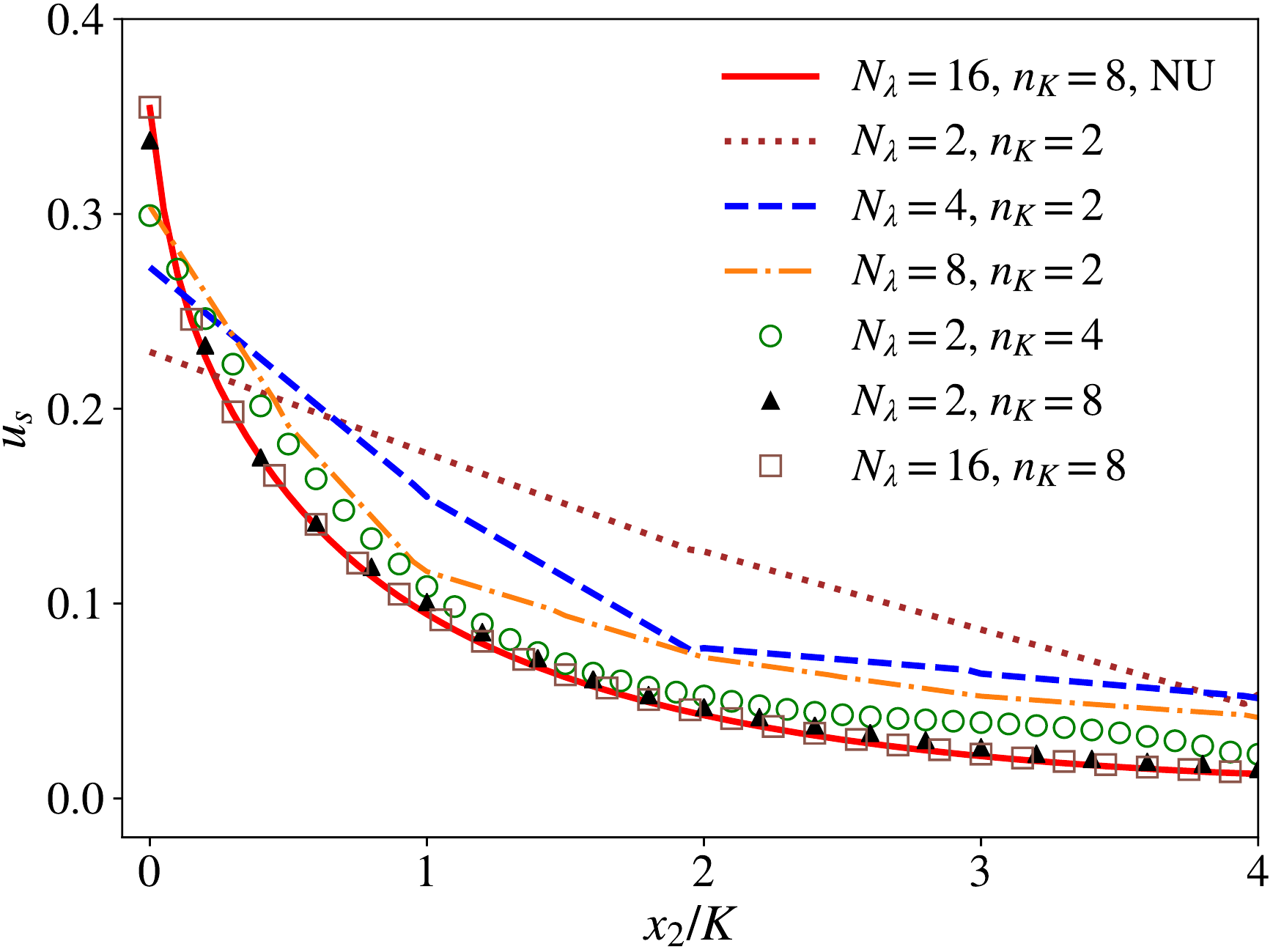}
	\caption{The Knudsen layer functions in  one-dimensional Couette flow obtained from DG schemes with different orders of accuracy $n_\text{K}$ (for the kinetic equation) and numbers of uniform cells $N_\lambda$. The Knudsen number is $\sqrt{\pi}K/2=0.01$. The order of DG scheme for macroscopic equations is $n_\text{S}=8$, and the bulk region is partitioned by 32 uniform cells. Note that the reference result (red solid line labeled with `NU') is obtained on the grid with 16 nonuniform cells in the Knudsen layer with refinement approaching to the wall. }
	\label{fig:Couette}
\end{figure}

Figure~\ref{fig:Couette} illustrates the Knudsen layer functions obtained from DG schemes with different values of $n_\text{K}$ and $N_\lambda$. The reference result is obtained when 16 nonuniform cells in the Knudsen layer with refinement near the wall are used, where the minimum and maximum cell sizes are 0.0089 and 0.46 mean free path, respectively; this nonuniform grid is more suitable to recover the divergence of velocity gradient at the wall~\cite{Jiang2016JCP}. It is concluded that using very few cells and lower order scheme cannot predict the correct Knudsen layer function. Accurate solution can only be obtained by increasing the order of the scheme and/or the number of cells in the Knudsen layer.

Table~\ref{Tab:Couette} shows the number of iteration steps, the defect velocity at the wall and the velocity gradient $k_1$ for different combinations of $N_\lambda$ and $n_\text{K}$, when $\epsilon$ defined in Eq.~\eqref{general_epsilon} is less than $10^{-5}$. The convergence can be reached around 30 iteration steps when the cell size in the Knudsen layer $L_\lambda$ is not smaller than 0.25 mean free path. However, as the cell size further reduces, the number of iteration increases. This is due to the fact that the local Knudsen number $Kn_\text{loc}$ appearing in Eq.~\eqref{Kn_loc} that depends on cell size becomes large, and the coefficient $\beta$ in Eq.~\eqref{GSIS_K30} for the update of macroscopic quantities  is less than one. Therefore, macroscopic quantities are not fully corrected by solutions from synthetic equations, which reduces the efficiency of GSIS a bit.

It is interesting to note that even when the Knudsen layer is not resolved, the velocity gradients~\eqref{velocity_gradient} in Table~\ref{Tab:Couette} all approach the reference value. This is understandable since according to Eq.~\eqref{KLF} the error introduced by inaccurate capture of Knudsen layer is $O(K)$, which is small compared to the wall velocity $O(1)$ in the near-continuum flow regimes. However, this case does not indicate that the capture of Knudsen layer is not important. In thermal effects where the velocity slip induced in the Knudsen layer is the origin of gas motion~\cite{Sone2002Book}, the Knudsen layer must be resolved, in order to obtain accurate flow filed in the whole computational domain, see the spatial discretization in Ref.~\cite{Taguchi2012JFM} for an example. In the following case, we will show evidence for this argument.


\begin{table}
	\caption{The number of iterative steps (Itr),  the defect velocity $u_s$ at the wall, and the Knudsen layer function $k_1={\mathrm{d}u_1(x_2=0.5)}/{\mathrm{d}x_2}$ under different combinations of $N_{\lambda}$ and $n_\text{K}$, in one-dimensional Couette flow with $\sqrt{\pi}K/2=0.01$. $L_{\lambda}=0.04/N_{\lambda}$ is the uniform cell size in the Knudsen layer. The bulk region is partitioned by $32$ uniform triangles and the order of DG scheme for macroscopic equations is $n_\text{S}=8$.}
	
	\centering
	\begin{tabular}{ccccccccccc}
		\hline
		\multirow{2}{*}{$N_{\lambda}$} & \multirow{2}{*}{$\frac{L_{\lambda}}{\sqrt{\pi}K/2}$} &  \multicolumn{3}{c}{$n_\text{K}=2$} & \multicolumn{3}{c}{$n_\text{K}=4$} & \multicolumn{3}{c}{$n_\text{K}=8$} \\
		\cline{3-11}
		& &  Itr & $u_s$ & $k_1$ & Itr & $u_s$ & $k_1$ & Itr & $u_s$ & $k_1$\\
		\hline
		2  & 2.00 & 43 & 0.229 & 1.957 & 27 & 0.299 & 1.955 & 27 & 0.338 & 1.955 \\
		4  & 1.00 & 32 & 0.273 & 1.957 & 26 & 0.322 & 1.955 & 30 & 0.349 & 1.955\\
		8  & 0.50 & 28 & 0.304 & 1.956 & 26 & 0.338 & 1.955 & 32 & 0.354 & 1.955\\
		16 & 0.25 & 36 & 0.324 & 1.956 & 36 & 0.349 & 1.955 & 43 & 0.355 & 1.955\\
		32 & 0.13 & 49 & 0.339 & 1.956 & 49 & 0.353 & 1.955 & 55 & 0.354 & 1.955\\
		64 & 0.06 & 62 & 0.348 & 1.956 & 48 & 0.354 & 1.955 & 66 & 0.353 & 1.955\\
		\hline
	\end{tabular}
	\label{Tab:Couette}
\end{table}

\subsection{Two-dimensional thermally-induced flow}\label{thermal_edge}
\begin{figure}
  \centering
  \includegraphics[trim=65 72 60 80, clip,scale=0.51]{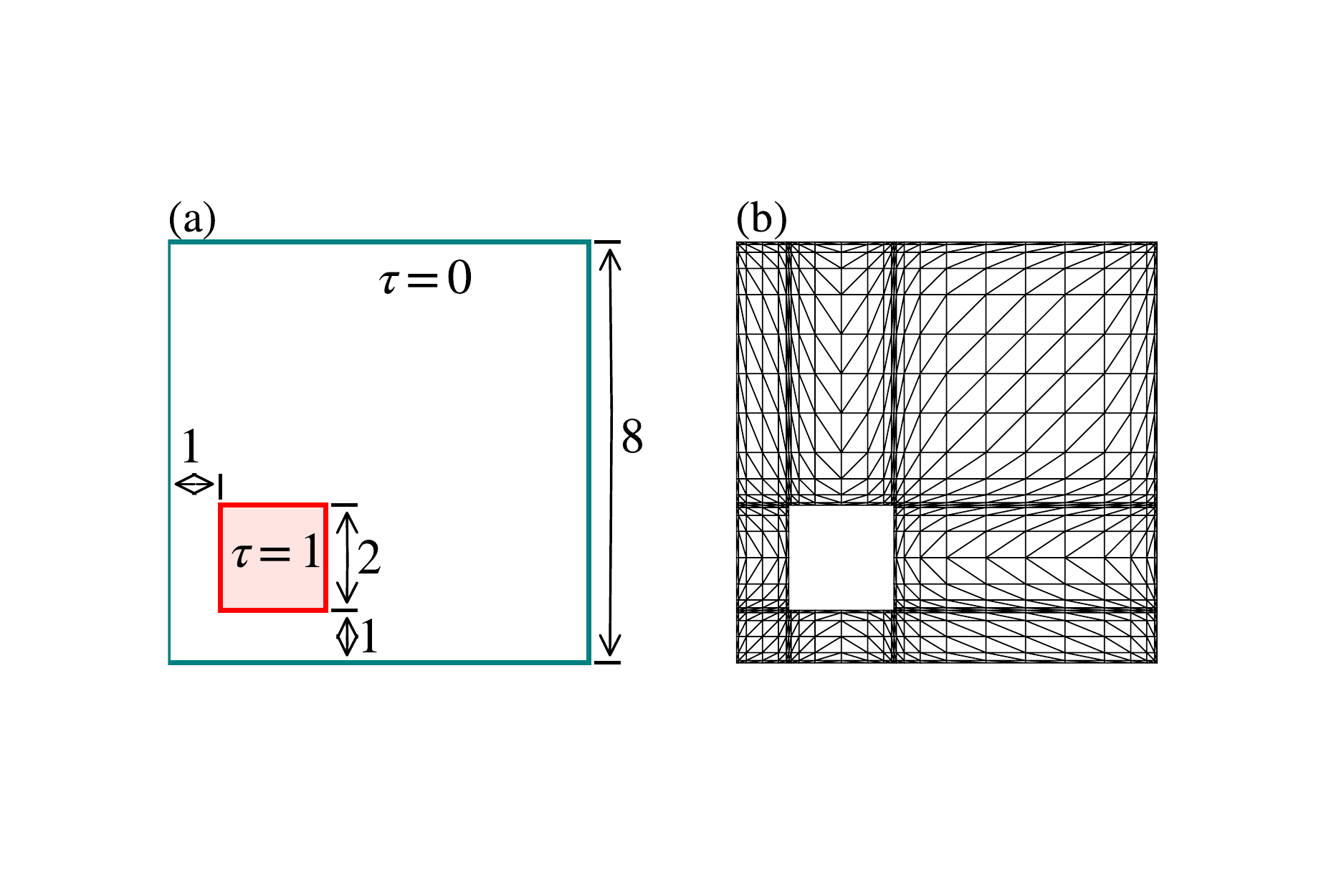}\ \includegraphics[trim=20 170 20 170,clip,scale=0.22]{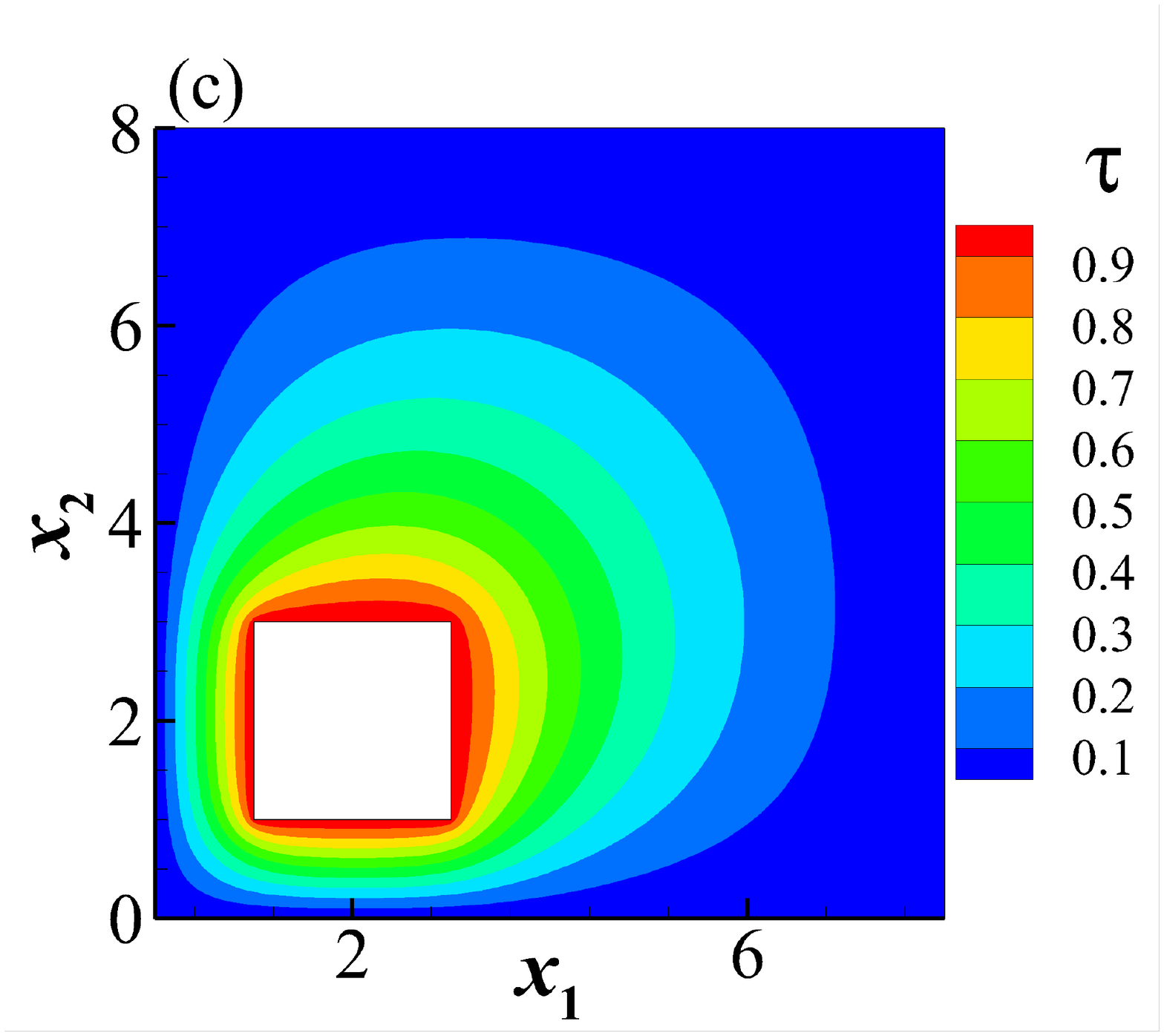}
  \caption{Two-dimensional thermal flow induced by squared hot beam encompassed in a squared cold chamber when $\sqrt{\pi}K/2=0.001$. Schematic illustrations of (a) geometry and (b) spatial discretization of structured triangles with refinement in the vicinity of solid walls. (c) typical temperature contours obtained by DG schemes of $n_\text{K}=n_\text{S}=6$ and on spatial mesh of 16104 triangles. Note that all the flow properties are symmetric about the diagonal joining the lower-left and upper-right corners of the chamber. }
  \label{fig:Beam}
\end{figure}

As depicted in Figure~\ref{fig:Beam}(a), we consider a two-dimensional flow induced by a hot beam that is encompassed in a cold chamber. Both the beam and chamber are square, with dimensions of $2\times2$ and $8\times8$, respectively. The beam with a (deviated) temperature of $\tau=1$ is placed with distance of 1 away from the left and bottom walls of the enclosure. The temperature of the chamber is $\tau=0$ and gas is filled between the beam and chamber. It is well known that no bulk flow is generated at Navier-Stokes limit and the gas temperature is governed by Fourier's law of heat conduction. However, under the non-equilibrium circumstances, thermal stress and thermal edge flows are induced by the rarefaction effects in the Knudsen layer~\cite{Sone2002Book}. 

We calculate the thermal flow at a challenge Knudsen number of $\sqrt{\pi}K/2=0.001$, when the characteristic flow length is chosen as the gap between the beam and chamber. In addition to the fact that the physical viscosity of gas is small, the magnitude of bulk velocity is also small; in this case high resolutions of both  Knudsen layer and bulk region are necessary, otherwise, under-resolution of Knudsen layer and/or large numerical dissipation will introduce error at the same or even larger order as that of the bulk velocity and lead to completely wrong result.

The computational domain is partitioned by structured triangles with refinement in the vicinity of walls, see Figure~\ref{fig:Beam}(b), which is characterized by the total number of triangles $N_\Delta$, the minimum cell size (the height of the local triangle) within Knudsen layer $L_\lambda$ and the maximum cell size in bulk region $L_\text{b}$. The diffuse boundary condition is imposed on the solid surfaces. The DG scheme is employed in the numerical simulation, and details to implement GSIS are the same as those in the previous subsections, except here we use $K_{th}=1$ for $n_\text{K}\leq4$ and $K_{th}=0.5$ for $n_\text{K}>4$. The typical temperature field is shown in Figure~\ref{fig:Beam}(c) that are obtained on the mesh of $N_\Delta=16104$, $L_\lambda=0.16K$ and $L_\text{b}=128K$ by DG schemes with $n_\text{K}=n_\text{S}=6$.

\begin{figure}
	\centering
	\includegraphics[trim=50 200 50 200, clip,scale=0.22]{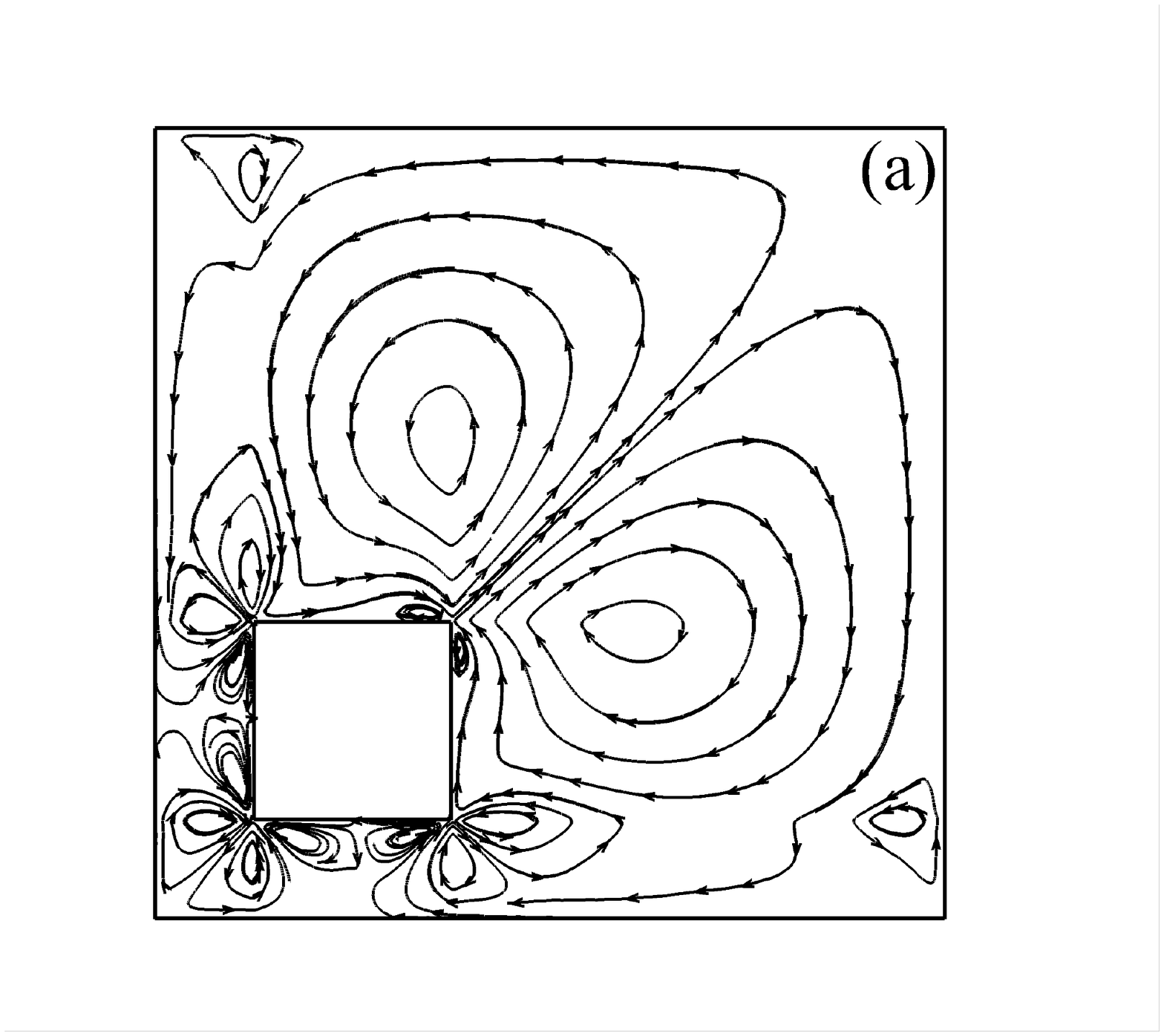}\includegraphics[trim=50 200 50 200,clip,scale=0.22]{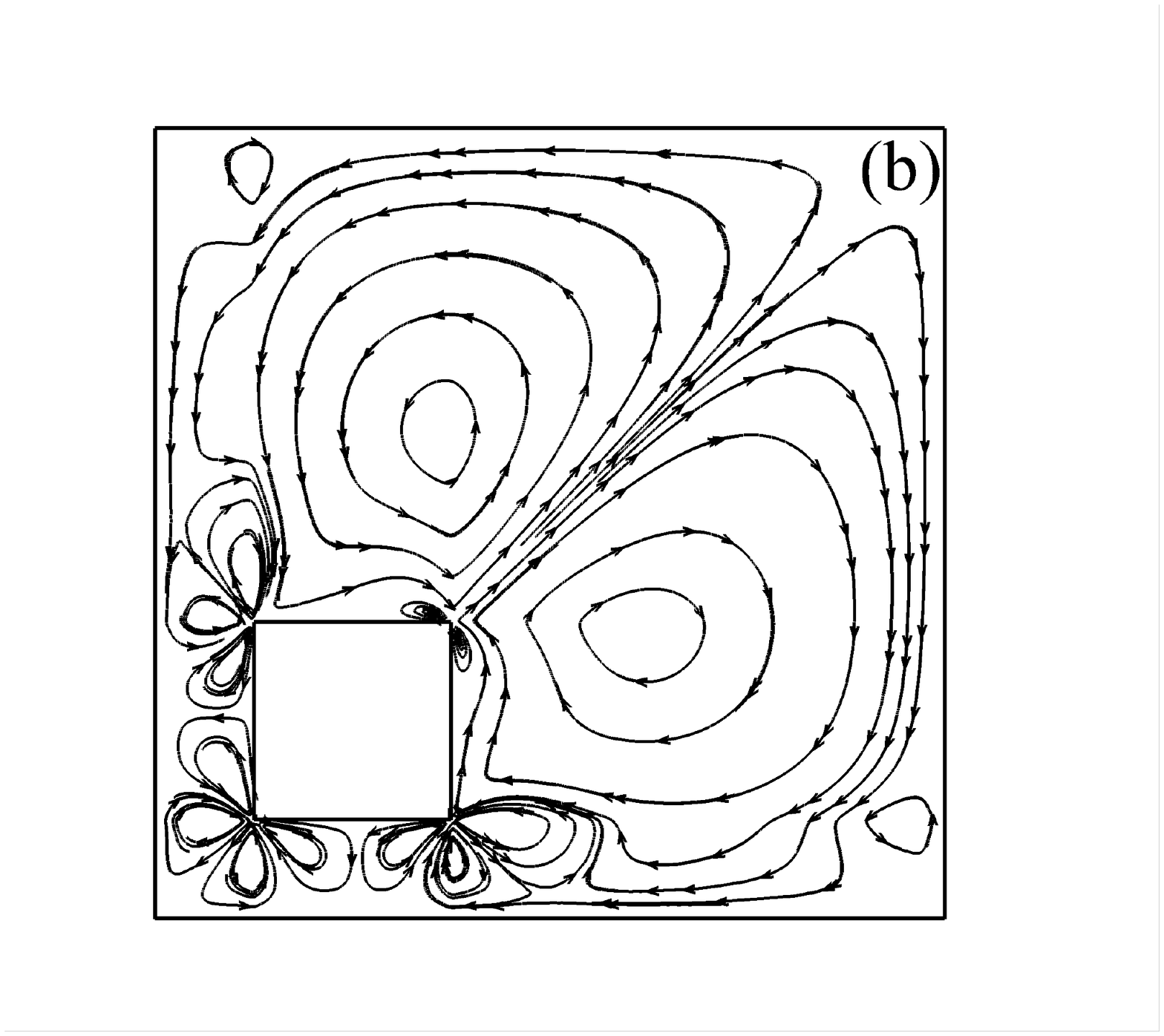}\includegraphics[trim=50 200 50 200,clip,scale=0.22]{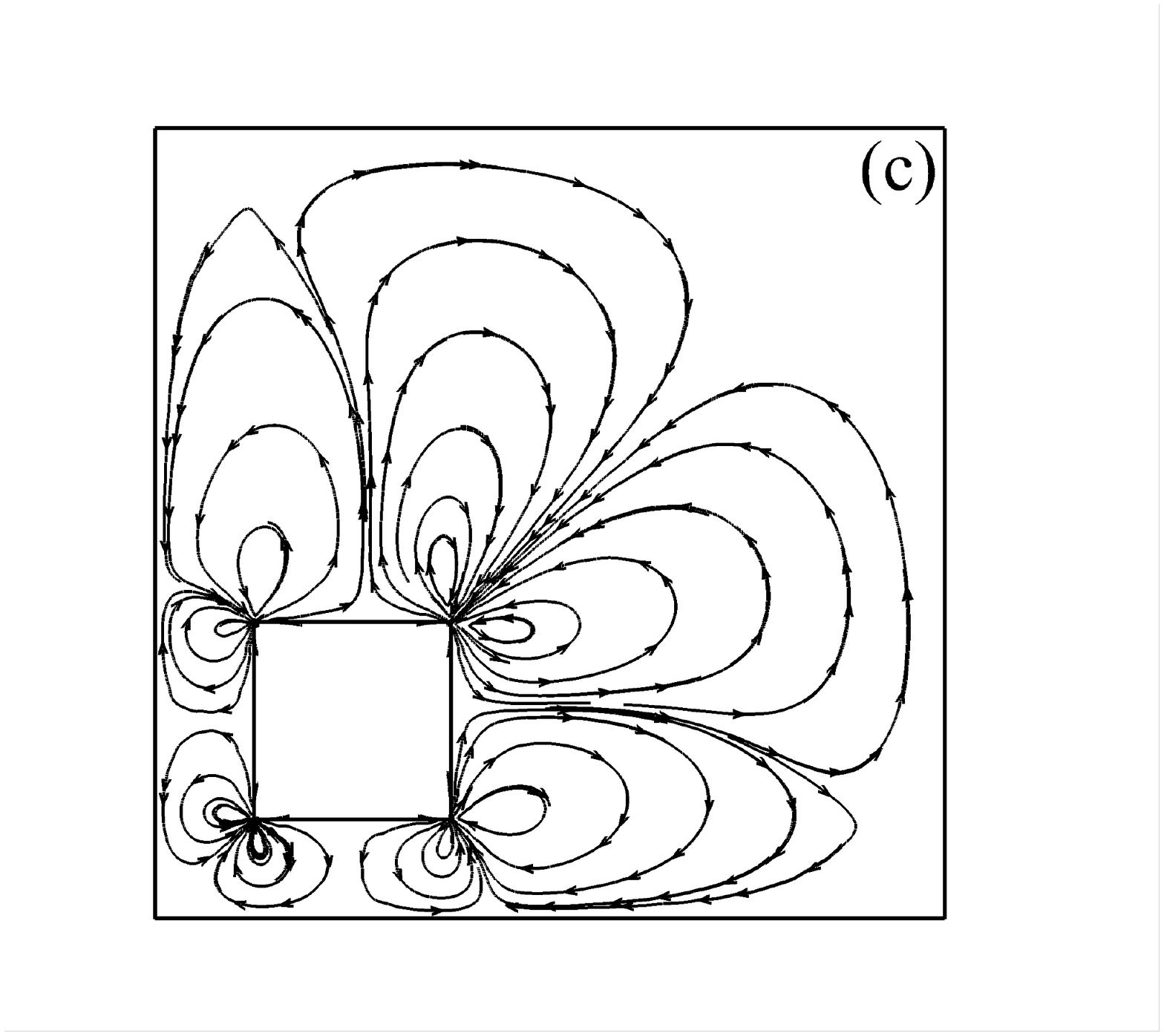}\\
	\includegraphics[trim=50 200 50 200,clip,scale=0.22]{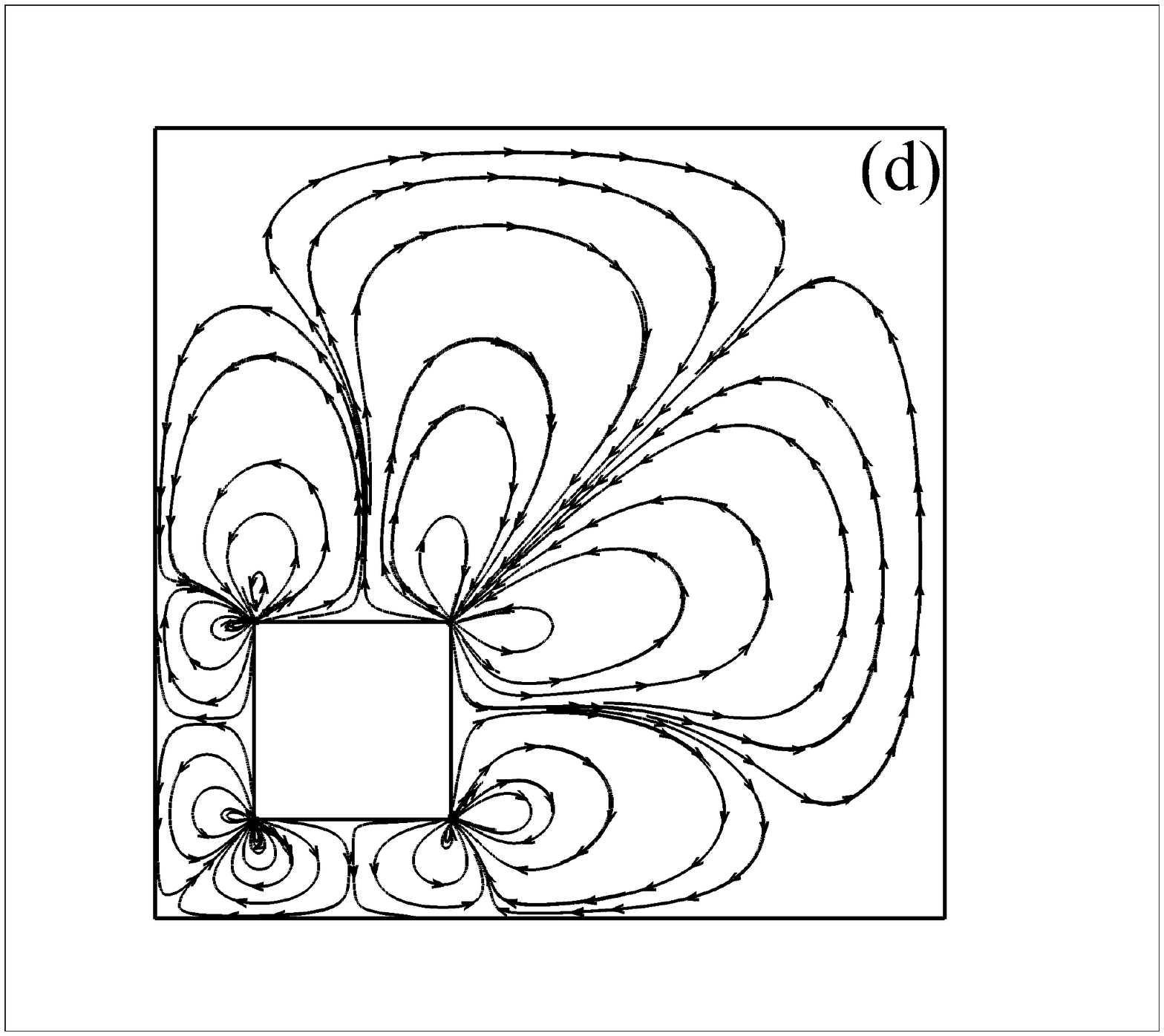}\includegraphics[trim=50 200 50 200,clip,scale=0.22]{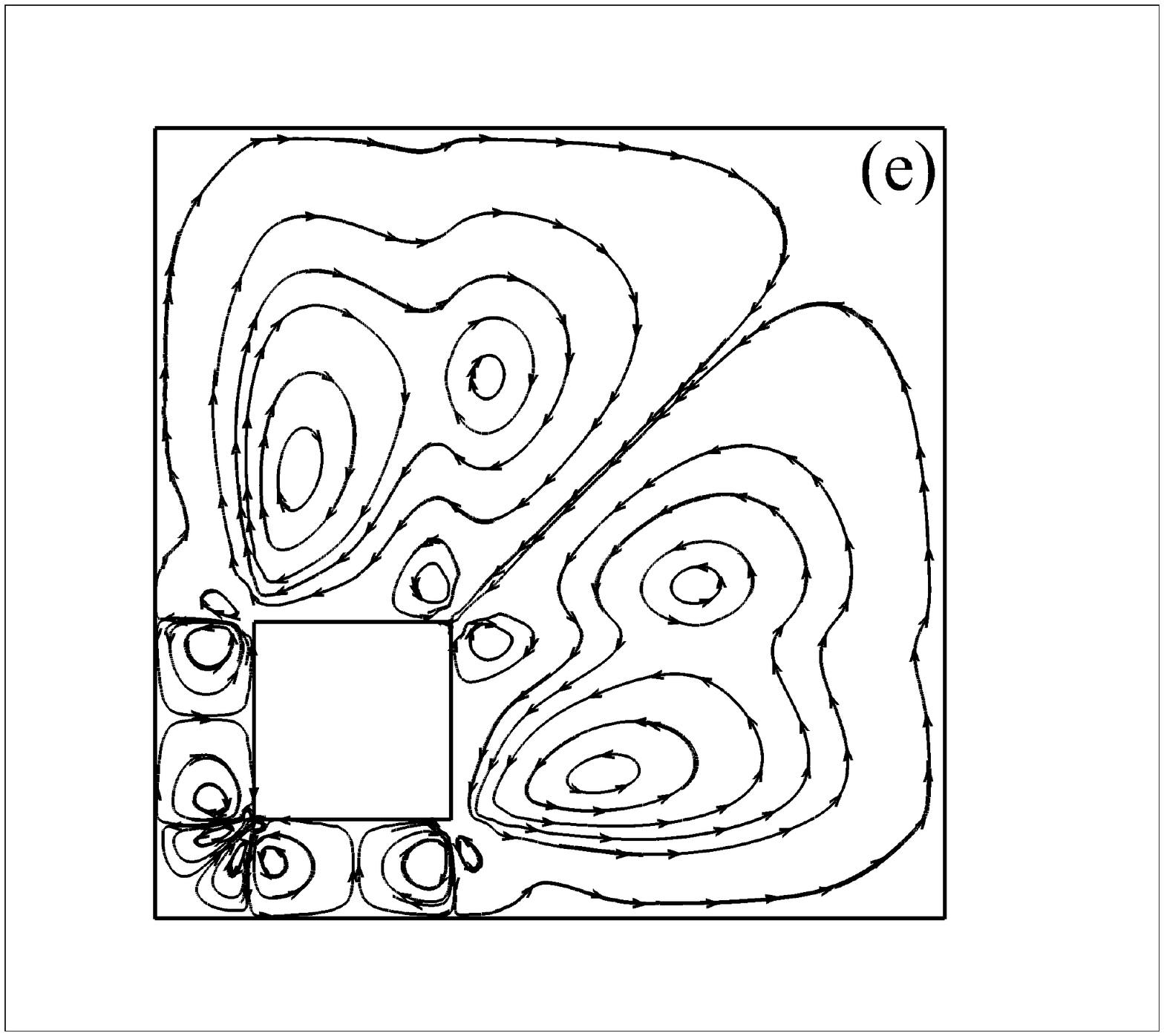}\includegraphics[trim=50 200 50 200,clip,scale=0.22]{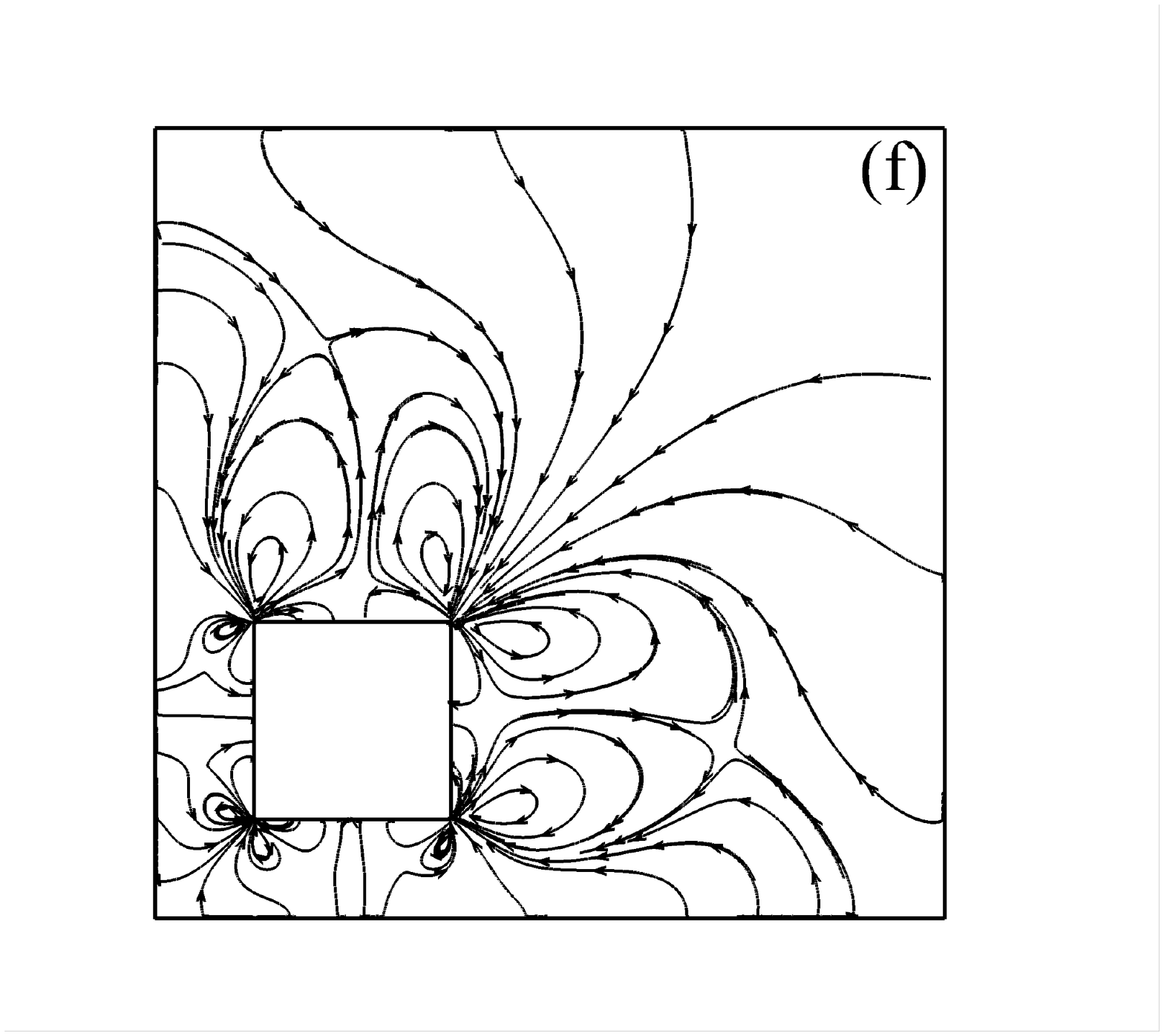}
	\caption{Streamlines obtained from DG schemes with different orders of accuracy and spatial meshes, in a two-dimensional thermal flow when $\sqrt{\pi}K/2=0.001$. (a) $L_\lambda=0.08K$, $L_\text{b}=128K$, and $n_\text{K}=n_\text{S}=2$;  (b) $L_\lambda=0.04K$, $L_\text{b}=64K$, and $n_\text{K}=n_\text{S}=2$; 
	(c) $L_\lambda=0.32K$, $L_\text{b}=256K$, and $n_\text{K}=n_\text{S}=4$; 
	(d) $L_\lambda=0.08K$, $L_\text{b}=128K$, and $n_\text{K}=n_\text{S}=6$; (e) $L_\lambda=0.08K$, $L_\text{b}=128K$, $n_\text{K}=2$ and $n_\text{S}=6$; (f) $L_\lambda=1.28K$, $L_\text{b}=64K$, and $n_\text{K}=n_\text{S}=4$.}
	\label{fig:BeamStream}
\end{figure}

We first consider the flow pattern  in Figure~\ref{fig:BeamStream}, which are obtained by the DG scheme of different orders and on different spatial meshes. On the same spatial mesh with cell size of 0.08 mean free path in the Knudsen layer and 128 mean free path in the bulk region, completely different flow patterns are obtained by second-order and sixth-order DG schemes, see Figure~\ref{fig:BeamStream}(a) and (d). In the flow field predicted by the second-order scheme, four small vortices are generated around the lower-left corner of the beam; while another four are developed near the upper-right corner of the beam. At each of the lower-right and upper-left corners of the beam, three more vortices appear. Two small vortices are observed near the lower-right and upper-left corners of the chamber. On the other hand, from the sixth-order DG scheme, only eight vortices are developed, with each corner of the beam having two. Therefore, on this spatial grid ($L_\lambda=0.08K$ and $L_\text{b}=128K$), the second-order scheme produces larger errors; reducing the cell size by half is even not adequate, see Figure~\ref{fig:BeamStream}(b). It is also worth noticing that only increasing the order of DG scheme for synthetic equations, the predicted flow field is still quiet different from that when the mesoscopic and macroscopic equations are all solved using sixth-order DG scheme, see flow field in Figure~\ref{fig:BeamStream}(e) obtained on mesh of $L_\lambda=0.08K$ and $L_\text{b}=128K$ and schemes of $n_\text{K}=2$ and $n_\text{S}=6$. These results show the importance of resolving the Knudsen layer, where the non-equilibrium effects are the sources of gas motion.

To further demonstrate the importance of resolving the Knudsen layer, we use the fourth-order DG scheme to solve the kinetic and synthetic equations on two different meshes: Figure~\ref{fig:BeamStream}(c) is obtained when  $L_\lambda=0.32K$ and $L_\text{b}=256K$, while Figure~\ref{fig:BeamStream}(f) is obtained when $L_\lambda=1.28K$ and $L_\text{b}=64K$. Figure~\ref{fig:BeamStream}(c) shows that, on the mesh where the Knudsen layer is resolved by relatively fine cells, the flow pattern with eight vortices are developed, which are very similar to the high resolution results in Figure~\ref{fig:BeamStream}(d), despite that the cell size in the bulk region is relatively large. However, when the bulk region is partitioned by fine cells but the Knudsen layer is under resolved by coarse cells, Figure~\ref{fig:BeamStream}(f) shows that although eight vortices can be observed, large errors appear near the walls, where gas seems to penetrate the solid walls.

\begin{figure}
  \centering
  \includegraphics[scale=0.4]{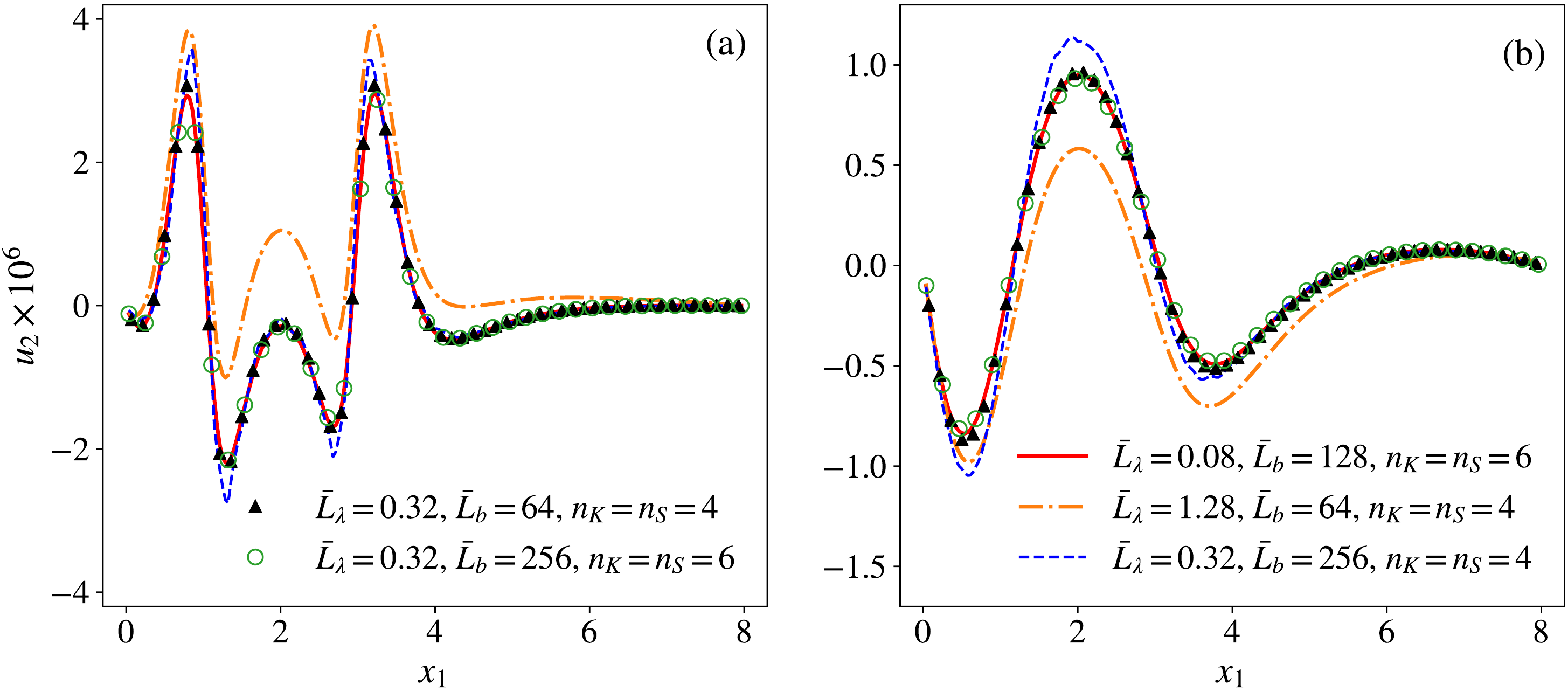}
  \caption{Flow velocities obtained from DG schemes with different orders of accuracy and meshes with different cell sizes, in a two-dimensional thermal flow when $\sqrt{\pi}K/2=0.001$. (a) vertical velocity $u_2$ along the horizontal line at {$x_2=0.5$}; (b) vertical velocity $u_2$ along the horizontal at {$x_2=4.5$}. $\bar{L}_\lambda$ and $\bar{L}_\text{b}$ are the minimum cell size in Knudsen layer and maximum cell size in bulk region, respectively, which are normalized by the mean free path of gas molecules, i.e. $\bar{L}_\lambda=L_\lambda/K$ and $\bar{L}_\text{b}=L_\text{b}/K$. The legends are applied to both figures.}
  \label{fig:BeamVelocity}
\end{figure}

\begin{table}
\caption{The number of iterative steps (Itr), the mean velocity magnitude $|\bar{\bm{u}}|_{x_1=0.5}$ along the horizontal line $x_2=0.5)$, as well as the computational time $t_\text{c}$ under different combinations of spatial meshes and schemes, in a two-dimensional thermal flow with $\sqrt{\pi}K/2=0.001$. $N_\Delta$ is the number of triangles in a mesh, $L_{\lambda}$ is the minimum cell size (triangle height) in Knudsen layer and $L_\text{b}$ is the maximum cell size in bulk region. `Err' is the relative error of the mean velocity magnitude $|\bm{u}|_{x_1=0.5}$ compared to the reference one obtained with $N_\Delta=20424$, $L_{\lambda}=0.08K$, $L_\text{b}=128K$ and $n_\text{K}=n_\text{S}=6$. The computational time $t_\text{c}$ is the wall time cost by each case that runs on 8 Intel Xeon-E5-2680 CPUs using OpenMP for parallelism. Note that the the molecular velocity space is discretized by $8\times8$-point Gauss-Hermite quadrature~\cite{SuWeiPRE2017}.}

\centering
\begin{tabular}{ccccccccc}
\hline
$N_{\Delta}$ & ${L_{\lambda}}/{K}$ & ${L_{b}}/{K}$ & $n_\text{K}$ & $n_\text{S}$ & $|\bar{\bm u}|\times10^6$ &  Err [\%] & Itr & $t_\text{c}$ [h] \\
 \hline
 8832  & 0.32 & 256 & 4 & 4 & 1.072 & 7.4 & 49 & 0.01\\
 16104 & 0.16 & 128 & 4 & 4 & 1.060 & 6.2 & 46 & 0.02\\
 24864 & 1.28 & 64  & 4 & 4 & 1.077 & 7.9 & 45 & 0.06\\
 30176 & 0.32 & 64  & 4 & 4 & 1.013 & 1.5 & 48 & 0.04 \\
 33024 & 0.16 & 64  & 4 & 4 & 1.013 & 1.5 & 46 & 0.04 \\
 8832  & 0.32 & 256 & 6 & 6 & 0.982 & 1.6 & 57 & 0.07 \\
 10400 & 0.16 & 256 & 6 & 6 & 0.991 & 0.7 & 55 & 0.12 \\
 16104 & 0.16 & 128 & 6 & 6 & 0.991 & 0.7 & 54 & 0.18 \\
 20424 & 0.08 & 128 & 6 & 6 & 0.998 & 0 & 60 & 0.21 \\
\hline
\end{tabular}
\label{Tab:Beam}
\end{table}

To quantitatively show the accuracy and efficiency of numerical simulations, we plot the vertical velocity along the horizontal lines $x_2=0.5$ and $x_2=4.5$ in Figure~\ref{fig:BeamVelocity}. As the refinement of the spatial mesh and increment of the order of accuracy of the DG schemes, the flow properties converge to the reference ones (red solid lines) obtained on the mesh of $L_\lambda=0.08K$ and $L_\text{b}=128K$ and by sixth-order DG scheme. 
In Table~\ref{Tab:Beam} we list the mean velocity along the horizontal line: $|\bar{\bm{u}}|=\frac{1}{8}\int^8_0|\bm{u}|\left(x_1,x_2=0.5\right)\mathrm{d}x_1$. The relative error between the mean velocity from the mesh of $L_\lambda=0.08K$ and $L_\text{b}=128K$ by sixth-order DG scheme and the one from the mesh of $L_\lambda=0.16K$ and $L_\text{b}=256K$ by the same order schemes is within 1\%, which indicates the accuracy of the reference result. The results for $|\bar{\bm{u}}|$ with different spatial discretizations and different orders of DG schemes clearly demonstrate the importance of resolving the Knudsen layer. The iteration steps and computational time cost by each case are also listed in Table~\ref{Tab:Beam}. For all the cases, GSIS can find the steady-state solution with 50 to 60 steps. The computational time to obtain accurate solution (with less than 2\% error in $|\bar{\bm{u}}|$ compared to the reference one) can be as little as several minutes. To the best of our knowledge, we are not aware of other numerical methods that are able to obtain accurate results within such a short time.

\section{Conclusions}\label{sec:conclusion}

In summary, based on the Fourier stability analysis, we have rigorously proven that the GSIS can lead to fast convergence of steady-state solutions. Due to the presence of continuity equation in macroscopic synthetic equations, the convergence property is different to other diffusion synthetic schemes for neutron transport and thermal radiation. Therefore, in GSIS the macroscopic quantities can only be partly updated according to the solution of synthetic equations when the Knudsen number is large. On the other hand, we have proven that the GSIS asymptotically preserves the Navier-Stokes limit when the spatial cell size is able to capture the physical solution. That is, in the bulk region the spatial cell size can be $\Delta{x}\sim{}O(1)$ as long as this size is adequate to capture the hydrodynamics. However, in the vicinity of solid walls the physical solution requires $\Delta{x}\sim{}O(K)$ in order to capture the Knudsen layer structures. This does not pose any problems to GSIS since it is an implicit solver so non-uniform spatial grid can be used. These analytical results have been confirmed by several numerical examples in the present paper.

It should be emphasized that GSIS bears the advantage that the mesoscopic kinetic equation and macroscopic synthetic equations can be solved by different schemes. Since most of the cost will be spent on the kinetic solver due to additional discretization in molecular velocity space,  in multi-scale simulation of rarefied gas dynamics, one can use high-order solver for synthetic equations, while low-order solver for kinetic equations, on a coarse spatial grid to save the computational time and cost,

\bibliographystyle{siamplain}
\bibliography{Bib}

\end{document}